\documentclass[%
twocolumn,
 amsmath,amssymb,
 aps,
 showpacs,
 longbibliography
]{revtex4-1}

\usepackage{amsmath}
\usepackage{mathrsfs}
\usepackage{graphicx}
\usepackage{dcolumn}
\usepackage{bm}
\usepackage{color}

\begin{document}

\preprint{APS/123-QED}

\title{
Exterior time scaling with the stiffness-free Lanczos time propagator: 
Formulation and application to atoms interacting with strong midinfrared lasers
}

\author{Haruhide Miyagi}
\author{Lars Bojer Madsen}
\affiliation{Department of Physics and Astronomy, Aarhus University, 8000 Aarhus C, Denmark}

\date{\today}

\begin{abstract}
Aiming at efficient numerical treatment of tunneling ionization of atoms and molecules by midinfrared (IR) lasers, exterior time-scaling (ETS) theory is formulated as a generalization of the time-scaled coordinate approach. The key idea of ETS is the division of the spatial volume into a small region around the nucleus and its outside; the radial coordinates are time scaled only in the outer region. The continuum components of photoelectron wave packets are prevented from reaching the edge of the spatial simulation volume, enabling the long-time evolution of wave packets with a relatively small number of basis functions without concerns of electron reflections. On the other hand, the bound-state components are free from shrinking toward the origin because of non-time scaling in the inner region. Hence, the equations of motion in ETS are less stiff than the ones in the original time-scaled coordinate approach in which the shrinking bound states make the equations of motion seriously stiff. For numerical implementation of ETS, the working equations are derived in terms of finite-element discrete-variable-representation functions. Furthermore, the stiffness-free Lanczos time propagator is introduced to remove any persistent stiffness in the treatment of mid-IR lasers due to the involvement of hundreds of angular momentum states. The test calculations for atomic hydrogen interacting with linearly polarized mid-IR pulses demonstrate the accuracy and numerical efficiency of this scheme, and exhibit its special capability if there is no recollision with the parent ion. Hence, ETS will show its true potential for the detailed analysis of photoelectron wave-packet dynamics in circularly or near-circularly polarized mid-IR fields. 
\end{abstract}

\pacs{32.80.Rm, 34.10.+x, 42.65.Ky}

\maketitle

\section{Introduction}

The recent advance of intense few-cycle light sources in the midinfrared (IR) region (wavelength: $\lambda\ge 3$ $\mu$m) is leading strong-field physics to a new direction~\cite{Wolter2015}. Its extreme nonlinear effect on matter generates high-order harmonic radiation covering the x-ray region~\cite{Popmintchev2012,Silva2015} with potentially narrowing the pulse width down to the zeptosecond~\cite{Garcia2013}, and is hence of very practical importance. Longer wavelength also ensures more detailed analysis and interpretation of experimental and numerical results, based on semiclassical (see, e.g., Refs.~\cite{Tate2007,Huismans2010,Huismans2012,Frolov2015}) and adiabatic~\cite{Tolstikhin2012,Ohmi2015} theories. Circularly or near-circularly polarized mid-IR pulses are ideal for precise attoclock measurements to elucidate tunneling dynamics~\cite{Eckle2008, Pfeiffer2012}, and for obtaining the information of molecular orbital structure from the photoelectron momentum distribution perpendicular to the polarization plane~\cite{Petersen2015}. Combing mid-IR lasers having different polarizations and colors may realize new means to reveal and control electron dynamics (see, e.g., recent references~\cite{Fleischer2014, Mancuso2015, Xie2015,Geng2015}). Numerical treatment of atoms and molecules in the mid-IR region, however, remains very challenging even within the single-active-electron (SAE) approximation~\cite{Kulander1992}. The reason comes not only from the involvement of many angular momentum states in the wave function, but also from the need to set the size of spatial volume proportional to $\lambda^2$, and also from the ponderomotive energy $U_p\propto\lambda^2$ which requires the use of denser grids or many basis functions as $\lambda$ increases~\cite{Pabst2013}. 

To manage the increasing numerical difficulty for large $\lambda$, the spatial volume is usually set as small as possible by employing a complex absorbing potential (CAP)~\cite{Riss1996} or exterior complex scaling (ECS)~\cite{McCurdy1991,He2007}, which prevents the high-energy continuum part of electron wave packets from reflection. This strategy is well suited for the analysis of low-energy photoelectrons and for computing dipole accelerations to investigate high-order-harmonic generation (HHG). An alternative strategy is to enlarge the spatial volume as time grows to prevent the reflection without losing the norm of wave function. Treatment of the time-dependent Schr\"odinger equation (TDSE) by periodic von Neumann basis with bi-orthogonal exchange~\cite{Takemoto2012} enables the extension of volume, which, however, results in a gradual increase of numerical cost. On the other hand, the time-scaled coordinate approach causes no such cost increase, and has, after its introduction in collision physics~\cite{Solovievt1985,Macek1999}, been developed aiming at efficient treatment of ionization by photoabsorption and electron impact~\cite{Sidky2000,Derbov2003,Roudnev2005,Serov2007,Serov2008,Hamido2011,Frapiccini2015}. After a long-time propagation under field-free conditions, the photoelectron wave packet in the scaled coordinate becomes stationary, from which the ionization cross section is extracted without projection onto the scattering wave function. This is advantageous in the treatment of many-electron systems, in particular, for the computation of the ndouble or multiple ionization cross section~\cite{Serov2007,Serov2008}. However, the time-scaled coordinate approach has a very serious shortcoming: The bound-state part of the electron wave packet shrinks toward the origin as time goes. Using dense grids or many basis functions around the origin to account for this shrinking makes the equations of motion stiff (see, e.g., Ref.~\cite{Press2007} for a discussion of stiffness of differential equations). In description of tunneling ionization at large $\lambda$, the equations become much stiffer and numerically untractable due to the involvement of hundreds of angular momentum states and the increase of the centrifugal potential barrier in the Hamiltonian. Although employing, e.g., Fatunla's method~\cite{Fatunla1978,Fatunla1980,Hamido2011,Frapiccini2015} or other elaborate time propagators may manage the problem, it is unfavorable to deploy many basis functions for the bound states since the research interest is in the description and analysis of the outgoing continuum part of the wave packet.

The aim of this paper is twofold: (i) formulation of exterior time-scaling (ETS) theory for extending numerical exploration toward the mid-IR region, and (ii) establishment of its stiffness-free numerical implementation. The ETS theory is a generalization of the original time-scaled coordinate approach which is hereafter, for comparison, referred to as global time scaling (GTS). The idea of ETS is to divide the spatial volume into two parts: a small domain around the nucleus, and its outside. The time scaling is carried out only to radial coordinates in the outer region. The continuum components in the outer region are hence prevented from reaching the edge of the spatial simulation volume, enabling the long-time evolution of the wave packet with a relatively small number of basis functions. In the inner region, on the other hand, the bound states are free from shrinking because of non-time scaling, and the equations of motion are expected to be less stiff than the ones in GTS. As many angular momentum states are involved, however, the equations in ETS inevitably become stiff. To address this problem, we propose a stiffness removal procedure which is in particular suitable for wave functions expanded in terms of finite-element discrete-variable-representation (FEDVR) functions (see, e.g., Ref.~\cite{Rescigno2000}) and time propagated by the Lanczos algorithm~\cite{Park1986}. This procedure is not specific to the ETS implementation but will also be applicable to a general class of equations appearing in atomic and molecular physics. Also note that, although in the following this paper aims at the treatment of one-electron atoms, the ETS method with the stiffness-free time propagation is applicable to the TD (restricted/generalized)-active-space configuration-interaction [TD-(RAS/GAS)CI] approach~\cite{Hochstuhl2012,Bauch2014} and may also be so to the $R$-matrix theory including time dependence (RMT)~\cite{Nikolopoulos2008,Moore2011a,Hart2014}, for instance, and can therefore be flexibly used to investigate many-electron atoms and molecules.

This paper is organized as follows. The ETS theory is formulated for atomic hydrogen in Sec.~\ref{Formulation}. Expanding the wave function in terms of FEDVR functions, Sec.~\ref{Practical} shows the derivation of the working equations for practical ETS implementation. Then Sec.~\ref{Stiffness-free Lanczos algorithm} is devoted to the analysis of stiffness and the discussion of its removal; the stiffness-free FEDVR-based Lanczos algorithm proposed in this section has a wide range of applicability. Based on the ETS theory with the stiffness-free procedure, Sec.~\ref{Ionization} demonstrates the tunneling ionization of atomic hydrogen in linearly polarized mid-IR lasers, and shows the accuracy and numerical efficiency of ETS. Section~\ref{Conclusion} concludes this work and provides an outlook. Atomic units are used throughout unless otherwise stated.

\section{\label{Formulation}Formulation}

We consider atomic hydrogen prepared in the ground state which then starts interacting with a light pulse linearly polarized along the $z$ axis. This  simple case is considered just for notational simplicity in the formulation; the generalization to many-electron atoms and molecules in arbitrarily polarized light fields is, at least formally, straightforward in any coordinate system. Expanding the wave function in terms of spherical harmonics with magnetic quantum number $m=0$, 
\begin{eqnarray}
\Psi({\bm r},t)
=
\frac{1}{r}
\sum_{\ell}
\psi_{\ell}(r,t)
Y_{\ell0}(\Omega), 
\label{Expansion0}
\end{eqnarray}
the TDSE leads to a set of coupled equations for the radial functions:
\begin{eqnarray}
i\left(\frac{\partial}{\partial t}\right)_{\hspace{-0mm}r}
\psi_{\ell}(r,t)
&=&
\left[
-\frac{1}{2}\frac{\partial^2}{\partial r^2}
+V_{\ell}(r)
\right]
\psi_{\ell}(r,t)
\nonumber
\\
&&
\hspace{-15mm}
+\sum_{\ell'}W_{\ell\ell'}(t)\psi_{\ell'}(r,t),
\label{TDSE0}
\end{eqnarray}
where 
\begin{eqnarray}
V_{\ell}(r)
=
\frac{\ell(\ell+1)}{2r^2}
-
\frac{1}{r},
\label{atomic potential}
\end{eqnarray}
and the light-atom interaction operator is treated within the dipole approximation,
\begin{widetext}
\begin{eqnarray}
W_{\ell\ell'}(t)
=
\left\{
\begin{array}{lcccl}
W^{\rm L}_{\ell\ell'}(r,t)&=&g_{\ell\ell'}F(t) r, & & {\rm in\;length\;gauge} \\ \\
W^{\rm V}_{\ell\ell'}(r,\partial_r,t)&=&\displaystyle{-ig_{\ell\ell'}A(t)\left[\frac{\partial}{\partial r}+\frac{\ell'(\ell'+1)-\ell(\ell+1)}{2r}\right]},
& & {\rm in\;velocity\;gauge} \\
\end{array}
\right.
\label{light_attom_interaction}
\end{eqnarray}
\end{widetext}
with 
$
g_{\ell\ell'}=
\sqrt{4\pi/3}
\int Y^*_{\ell 0}(\Omega)Y_{10}(\Omega) Y_{\ell'0}(\Omega)d\Omega
$, and the vector potential $A(t)$ and the electric field $F(t)\big(= -dA(t)/dt\big)$ of light. The gauge-specific notation, $W^{\rm L}_{\ell\ell'}(r,t)$ and $W^{\rm V}_{\ell\ell'}(r,\partial_r,t)$, is in the following used only when their distinction is necessary. Note that $(\partial/\partial t)_x$ denotes the partial time derivative for a fixed value of $x(=r$ or $\xi$). Such an explicit notation is not needed in Eq.~\eqref{TDSE0} but in the following helps avoid unnecessary confusion.

The formulation of ETS commences with introducing a spherical surface, $\Sigma$, the radius of which is $r_{\Sigma}$. Setting the center of $\Sigma$ at the origin, the configuration space is divided into its inner and outer regions. Let $\Sigma$ itself belong to the inner region. We then define an ETS map by
\begin{eqnarray}
\xi(r,t)
=
\left\{
\begin{array}{lcc}
r, & & (0\le r\le r_{\Sigma}), \\ \\
r_{\Sigma}+(r-r_{\Sigma})/R(t),& & (r_{\Sigma}< r<\infty), \\
\end{array}
\right.
\label{ETS_map}
\end{eqnarray}
where $R(t)(\ge 1)$ is a smooth increasing function of time; its explicit form is given later [see Eq.~\eqref{R(t)_2} in Sec.~\ref{Ionization}]. Noting the mutual dependence between $r$ and $\xi$, i.e., $r=r(\xi,t)$ and $\xi=\xi(r,t)$, the differentiation of Eq.~\eqref{ETS_map} reads $dr(\xi,t)=d\xi(r,t)$ for $0\le r\le r_{\Sigma}$, and $dr(\xi,t)=\big[\xi(r,t)-r_{\Sigma}\big] dR(t) + R(t)d\xi(r,t)$ for $r_{\Sigma}< r<\infty$. Hence the differential equation, $dr(\xi,t)=0$, leads to relations:
\begin{eqnarray}
\left(\frac{\partial}{\partial t}\right)_{\hspace{-0mm}r}
=
\left\{
\begin{array}{lcc}
{\displaystyle \left(\frac{\partial}{\partial t}\right)_{\hspace{-0mm}\xi}}, & & (0\le r\le r_{\Sigma}) \\ \\
{\displaystyle \left(\frac{\partial}{\partial t}\right)_{\hspace{-0mm}\xi}
-
\frac{\dot{R}(t)}{R(t)}(\xi-r_{\Sigma})\frac{\partial}{\partial \xi}},& & (r_{\Sigma}< r<\infty) \\
\end{array}
\right..
\nonumber \\
\label{partial-derivative2}
\end{eqnarray}
In $0\le r\le r_{\Sigma}$, Eq.~\eqref{ETS_map} is just an identity mapping. The coupled equations obeyed by the radial functions in $0\le \xi\le r_{\Sigma}$ are thus obtained from Eq.~\eqref{TDSE0} by simply replacing $r$ by $\xi$:
\begin{widetext}
\begin{eqnarray}
i\left(\frac{\partial}{\partial t}\right)_{\hspace{-0mm}\xi}
\psi_{\ell}(\xi,t)
=
\left[
-\frac{1}{2}\frac{\partial^2}{\partial \xi^2}
+V_{\ell}(\xi)
\right]
\psi_{\ell}(\xi,t)
+
\sum_{\ell'}
W_{\ell\ell'}(t)
\psi_{\ell'}(\xi,t),
\hspace{3mm}
(0\le \xi\le r_{\Sigma}).
\label{TDSE_global_TSCSP_in}
\end{eqnarray}
In $r_{\Sigma}< r<\infty$, on the other hand, using the second lines of Eqs.~\eqref{ETS_map} and \eqref{partial-derivative2} in Eq.~\eqref{TDSE0} leads to 
\begin{eqnarray}
&&i\left(\frac{\partial}{\partial t}\right)_{\hspace{-0mm}\xi}
\psi_{\ell}[r_{\Sigma}+R(t)(\xi-r_{\Sigma}),t]
=
\bigg[
-\frac{1}{2[R(t)]^2}
\frac{\partial^2}{\partial \xi^2}
+
V_{\ell}[r_{\Sigma}+R(t)(\xi-r_{\Sigma})\big]
+
i\frac{\dot{R}(t)}{R(t)}(\xi-r_{\Sigma})
\frac{\partial}{\partial \xi}
\bigg]
\psi_{\ell}[r_{\Sigma}+R(t)(\xi-r_{\Sigma}),t]
\nonumber
\\
&&
\hspace{55mm}
+
\sum_{\ell'}W_{\ell\ell'}(t)\psi_{\ell'}[r_{\Sigma}+R(t)(\xi-r_{\Sigma}),t],
\hspace{5mm}
(r_{\Sigma}< \xi<\infty).
\label{TDSE1_ETS}
\end{eqnarray}
Defining new radial functions by 
\begin{eqnarray}
\phi_{\ell}(\xi,t)
=
\sqrt{R(t)}\exp\Big[-iR(t)\dot{R}(t)(\xi-r_{\Sigma})^2/2\Big]
\psi_{\ell}(r,t),
\hspace{5mm}
(r_{\Sigma}< r<\infty),
\label{wave_function}
\end{eqnarray}
Eq.~\eqref{TDSE1_ETS} is recast into a concise form:
\begin{eqnarray}
&&
i\left(\frac{\partial}{\partial t}\right)_{\hspace{-0mm}\xi}
\phi_{\ell}(\xi,t)
=
\bigg[
-\frac{1}{2[R(t)]^2}
\frac{\partial^2}{\partial \xi^2}
+
V_{\ell}[r_{\Sigma}+R(t)(\xi-r_{\Sigma})\big]
+
\frac{R(t)\ddot{R}(t)}{2}(\xi-r_{\Sigma})^2
\bigg]
\phi_{\ell}(\xi,t)
+
\sum_{\ell'}\mathcal{W}_{\ell\ell'}(t)
\phi_{\ell'}(\xi,t),
\nonumber
\\
&&
\hspace{30mm}
(r_{\Sigma}<\xi<\infty),
\label{TDSE_global_TSCSP_out}
\end{eqnarray}
with
\begin{eqnarray}
\mathcal{W}_{\ell\ell'}(t)
=
\left\{
\begin{array}{lcl}
W^{\rm L}_{\ell\ell'}[r_{\Sigma}+R(t)(\xi-r_{\Sigma}),t\big] & & {\rm in\;length\;gauge}, \\ \\
W^{\rm V}_{\ell\ell'}[r_{\Sigma}+R(t)(\xi-r_{\Sigma}),\partial_{R(t)\xi},t\big]  + g_{\ell\ell'} A(t)\dot{R}(t)(\xi-r_{\Sigma})& & {\rm in\;velocity\;gauge}, \\
\end{array}
\right.
\end{eqnarray}
\end{widetext}
where $g_{\ell\ell'} A(t)\dot{R}(t)(\xi-r_{\Sigma})$ comes from the exponent in Eq.~\eqref{wave_function} (a corresponding factor, $A(t)\dot{R}(t)\xi$, is missing in Eq.~(8) of Refs.~\cite{Hamido2011,Frapiccini2015} where GTS is formulated for one-dimensional systems). Importantly, although Eqs.~\eqref{TDSE_global_TSCSP_in} and \eqref{TDSE_global_TSCSP_out} are defined in non-overlapping radial intervals, $0\le \xi\le r_{\Sigma}$ and $r_{\Sigma}< \xi<\infty$, respectively, and may seem to be independent of each other, the equations are indeed coupled by the requirement of continuous differentiability of the wave function across $\Sigma$: $\Psi(\bm{r},t)\big|_{r=r_{\Sigma}}=\Psi(\bm{r},t)\big|_{r=r_{\Sigma}+0}$ and $\partial\Psi(\bm{r},t)/\partial r\big|_{r=r_{\Sigma}}=\partial\Psi(\bm{r},t)/\partial r\big|_{r=r_{\Sigma}+0}$, i.e., for each $\ell$,
\begin{subequations}
\begin{eqnarray}
\psi_{\ell}(r_{\Sigma},t)
&=&
\frac{1}{\sqrt{R(t)}}
\phi_{\ell}(r_{\Sigma}+0,t),
\label{continuity0}
\\
\frac{\partial}{\partial\xi}
\psi_{\ell}(\xi,t)
\Big|_{\xi=r_{\Sigma}}
&=&
\frac{1}{[R(t)]^{3/2}}
\frac{\partial}{\partial\xi}
\phi_{\ell}(\xi,t)
\Big|_{\xi=r_{\Sigma}+0}.
\label{diff_continuity}
\end{eqnarray}
\end{subequations}

\begin{figure*}
\begin{center}
\begin{tabular}{c}
\resizebox{175mm}{!}{\includegraphics{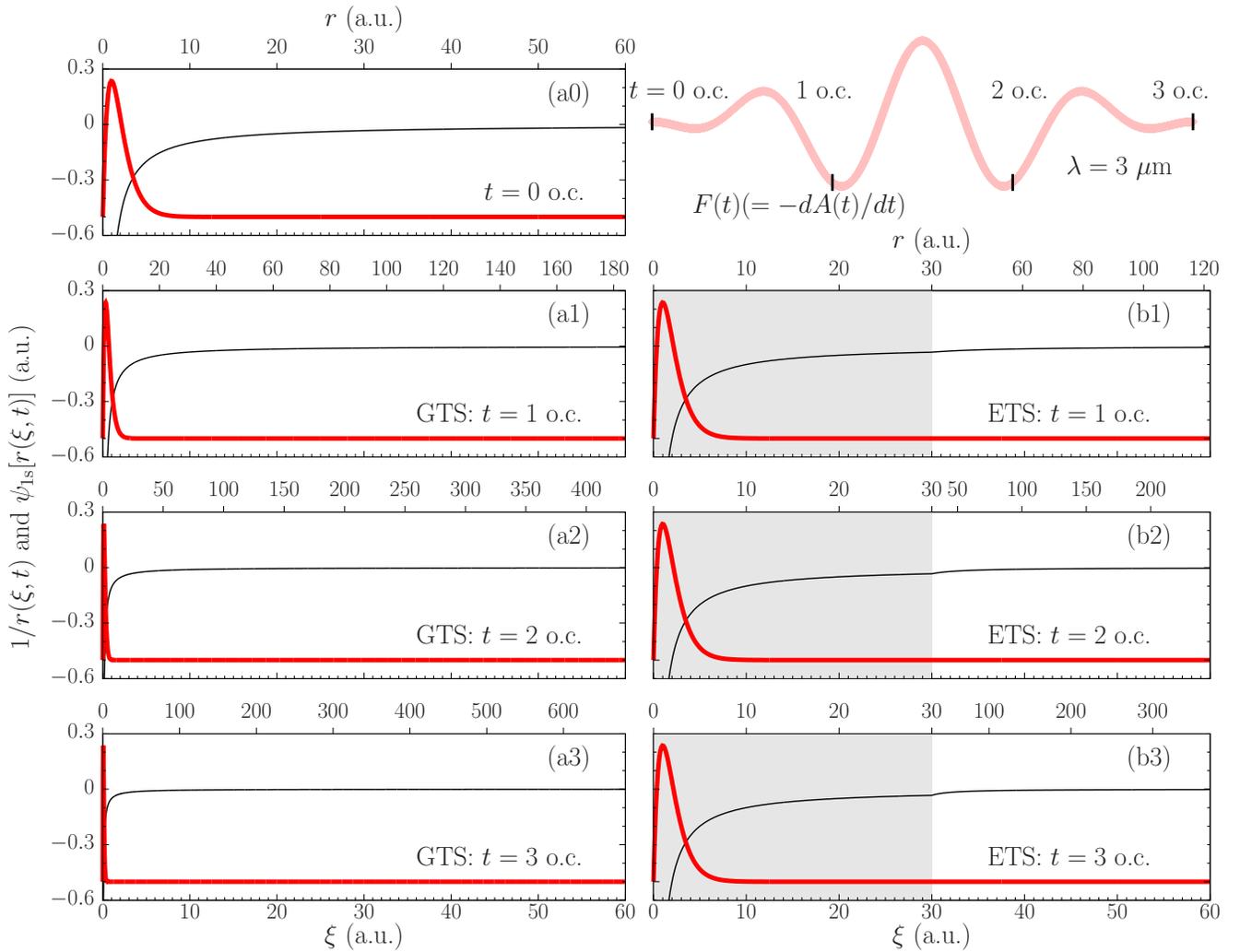}}
\end{tabular}
\caption{
\label{fig_bound_states}
(Color online) Illustration of the shrinking of the ground-state wave function in GTS and the non-shrinking in ETS. The Coulomb potential function, $1/r(\xi,t)$ (thin lines), and the radial part of the ground-state wave function of atomic hydrogen, $\psi_{\rm 1s}[r(\xi,t)]=2r(\xi,t)\exp[-r(\xi,t)]$ [thick (red) lines, shifted $-0.5$ downward], are plotted at zeroth, first, second, and third optical cycles (o.c.), i.e., at $t=(2\pi/\omega)\times n$ ($n=0,1,2$, and $3$). The temporal unit is measured in terms of one optical cycle of a mid-IR laser with $\lambda=3$ $\mu$m $=5.669\times10^4$. The plot is made under laser-free conditions, but, in connection with the numerical demonstration later, the top-right corner displays the electric field $F(t)=(-dA(t)/dt)$ of the same three-cycle pulse used in Sec.~\ref{HHG}. The time-scaling factor $R(t)$ is specified by Eq.~\eqref{R(t)_2} with $R_{\infty}=0.01$, which is also the same as in Sec.~\ref{HHG}. The value of $r_{\Sigma}$ is set to sero in GTS [(a1--3)], and $30$ in ETS [(b1--3)]. The shaded region in (b1--3) indicates the radial interval of the inner region ($0\le r \le r_{\Sigma}$) where the bound states are free from shrinking because of non-time scaling. The top and bottom of each panel represent the $r\big(=r(\xi,t)\big)$ and $\xi$ coordinates, respectively.
}
\end{center}
\end{figure*}

It is readily seen that ETS is a generalized concept of GTS; setting $r_{\Sigma}=0$ reduces the ETS map [Eq.~\eqref{ETS_map}] to the original GTS map, $\xi(r,t)=r/R(t)$; Eq.~\eqref{TDSE_global_TSCSP_out} becomes, e.g., Eq.~(8) in Refs.~\cite{Hamido2011,Frapiccini2015} except for the missing factor in the velocity gauge. As in the equations of motion in the conventional GTS, Eq.~\eqref{TDSE_global_TSCSP_out} indicates the introduction of an effective nucleus charge, $1/R(t)$, an effective electron mass, $[R(t)]^2$, and a temporal harmonic potential only while $\ddot{R}(t)>0$, preventing the electron from escaping to infinity. An important difference of ETS from GTS is that ETS avoids the complication of the shrinking of the bound states toward the origin present in GTS because of the time scaling only in the outer region. Figure~\ref{fig_bound_states} gives the comparison by illustrating the Coulomb potential function, $1/r(\xi,t)$, and the radial part of the ground-state wave function of atomic hydrogen, $\psi_{\rm 1s}[r(\xi,t)]=2r(\xi,t)\exp[-r(\xi,t)]$. The plot is made under laser-free conditions, but, supposing the application to mid-IR lasers with $\lambda=3$ $\mu$m $=5.669\times10^4$ (see Sec.~\ref{HHG}), the temporal unit is measured in its optical cycle (o.c.), $2\pi/\omega=413.7$, where $\omega=2\pi/\alpha\lambda=0.01519$ with the fine-structure constant $\alpha=1/137.036$. The time-scaling factor $R(t)$ is specified by Eq.~\eqref{R(t)_2} with $R_{\infty}=0.01$ (see the discussion in Sec.~\ref{Ionization}). While the Coulomb potential function and the radial function in GTS shrink toward the origin in the $\xi$ coordinate [Figs.~\ref{fig_bound_states}(a1--3)], they remain almost unchanged in ETS [Figs.~\ref{fig_bound_states}(b1--3)] even across $r_{\Sigma}=30$.

\section{\label{Practical}Practical FEDVR-based formalism}

The numerical implementation of ETS requires the expansion of the radial functions, $\psi_{\ell}(\xi,t)$ and $\phi_{\ell}(\xi,t)$ of Eqs.~\eqref{TDSE_global_TSCSP_in} and \eqref{TDSE_global_TSCSP_out}, in terms of a set of basis functions. There may be several options: grids, B splines, a variety of DVR functions, and combinations of them in the inner and outer regions. Among the various possibilities, we choose the FEDVR functions and derive the working equations for the ETS implementation.

\begin{figure*}
\begin{center}
\begin{tabular}{c}
\resizebox{175mm}{!}{\includegraphics{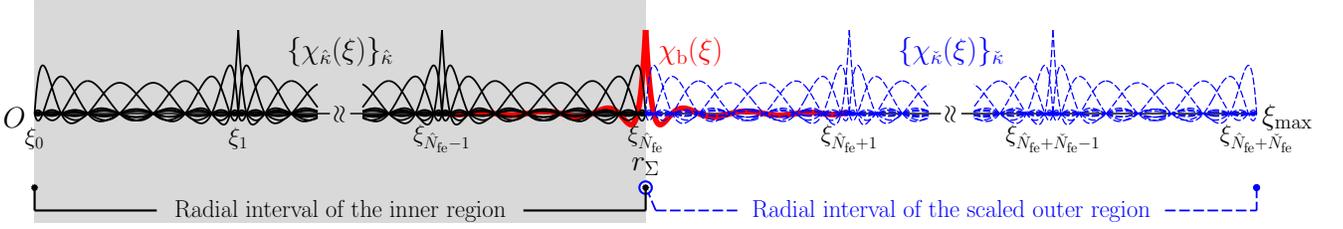}}
\end{tabular}
\caption{
\label{fedvr_basis}
(Color online) Setup of the FEDVR functions in the $\xi$ coordinate. Dividing the interval $[0,\xi_{\max}]$ into two parts, $[0,r_{\Sigma}]$ (shaded area) and $(r_{\Sigma},\xi_{\max}]$, they are further subdivided into $\hat{N}_{\rm fe}$ and $\check{N}_{\rm fe}$ intervals, respectively, having the same width $\Delta\xi$. Defining a set of Lobatto DVR functions with $N_{\rm dvr}$ quadrature points in each FE, the FEDVR functions are constructed along the standard prescription (cf. Ref.~\cite{Rescigno2000}). The FEDVR functions are classified into three categories, $\big\{\chi_{\hat{\kappa}}(\xi)|{\hat{\kappa}=1},\cdots,{(N_{\rm dvr}-1)\hat{N}_{\rm fe}-1}\big\}$, $\big\{\chi_{\rm b}(\xi)\big(\equiv\chi_{(N_{\rm dvr}-1)\hat{N}_{\rm fe}}(\xi)\big)\big\}$, and $\big\{\chi_{\check{\kappa}}(\xi)|{\check{\kappa}=(N_{\rm dvr}-1)\hat{N}_{\rm fe}+1},\cdots,{(N_{\rm dvr}-1)(\hat{N}_{\rm fe}+\check{N}_{\rm fe})-1}\big\}$, and illustrated by solid, thick solid (red), and dashed (blue) lines, respectively. Although the domain of every FEDVR function extends over the whole interval, $[0,\xi_{\max}]$, each function is explicitly illustrated only in its non-vanishing region. 
}
\end{center}
\end{figure*}

\subsection{\label{FEDVR function-setup}FEDVR functions}

Let a closed interval, $[0,\xi_{\max}]$, with $\xi_{\max}> r_{\Sigma}$, be the range of the scaled-radial coordinate, $\xi$. We then define a set of FEDVR functions over the interval as follows: Let the two intervals, $[0,r_{\Sigma}]$ and $(r_{\Sigma},\xi_{\max}]$, divided at $\xi=r_{\Sigma}$, be further subdivided into $\hat{N}_{\rm fe}$ and $\check{N}_{\rm fe}$ intervals (FEs), respectively:
\begin{subequations}
\begin{eqnarray}
\hspace{-7mm}
{[}0,r_{\Sigma}]
&=&
\big[\xi_{0},\xi_{1}\big]\cup\big(\xi_{1},\xi_{2}\big]\cup\cdots\cup
\big(\xi_{\hat{N}_{\rm fe}-1},\xi_{\hat{N}_{\rm fe}}\big],
\\
\hspace{-7mm}
{(}r_{\Sigma},\xi_{\max}{]}
&=&
\big(\xi_{\hat{N}_{\rm fe}},\xi_{\hat{N}_{\rm fe}+1}\big]\cup\cdots
\nonumber
\\
&&
\hspace{5mm}
\cup
\big(\xi_{\hat{N}_{\rm fe}+\check{N}_{\rm fe}-2},\xi_{\hat{N}_{\rm fe}+\check{N}_{\rm fe}-1}\big]
\nonumber
\\
&&
\hspace{5mm}
\cup
\big(\xi_{\hat{N}_{\rm fe}+\check{N}_{\rm fe}-1},\xi_{\hat{N}_{\rm fe}+\check{N}_{\rm fe}}\big],
\end{eqnarray}
\end{subequations}
where $\xi_{0}=0$, $\xi_{\hat{N}_{\rm fe}}=r_{\Sigma}$, and $\xi_{\hat{N}_{\rm fe}+\check{N}_{\rm fe}}=\xi_{\max}$. Let every interval have the same width $\Delta \xi=\xi_{\max}/(\hat{N}_{\rm fe}+\check{N}_{\rm fe})$. Let us then define the Lobatto DVR functions with $N_{\rm dvr}$ quadrature points. Following the standard prescription (see, e.g., Ref.~\cite{Rescigno2000}), a set of the FEDVR functions, $\big\{\chi_{\kappa}(\xi)|{\kappa=1},\cdots,{(N_{\rm dvr}-1)(\hat{N}_{\rm fe}+\check{N}_{\rm fe})-1}\big\}$, is composed. In the construction, two Lobatto DVR functions having quadrature points at $\xi_0$ and $\xi_{\max}$ are removed; consequently, every FEDVR function is zero at both edges of the domain, i.e., $\chi_{\kappa}(0)=\chi_{\kappa}(\xi_{\max})=0$.

The FEDVR functions are then classified into three groups: $\big\{\chi_{\hat{\kappa}}(\xi)|{\hat{\kappa}=1},\cdots,{(N_{\rm dvr}-1)\hat{N}_{\rm fe}-1}\big\}$, $\big\{\chi_{\rm b}(\xi)\big(\equiv\chi_{(N_{\rm dvr}-1)\hat{N}_{\rm fe}}(\xi)\big)\big\}$, and $\big\{\chi_{\check{\kappa}}(\xi)|{\check{\kappa}=(N_{\rm dvr}-1)\hat{N}_{\rm fe}+1},\cdots,{(N_{\rm dvr}-1)(\hat{N}_{\rm fe}+\check{N}_{\rm fe})-1}\big\}$. Note that the FEDVR functions belonging to the first and third groups, respectively, are non-vanishing only below and above $r_{\Sigma}$, and are distinguished, if need be, by the accent symbols \textit{hat}, `$\hat{\hspace{3mm}}$', and \textit{check}, `$\check{\hspace{3mm}}$', atop the index $\kappa$. On the other hand, the second group consists only of $\chi_{\rm b}(r)$, which is the bridge function across $r_{\Sigma}$ (see, e.g., Ref.~\cite{Rescigno2000} for a discussion of the bridge function). Figure~\ref{fedvr_basis} illustrates the setup of the FEDVR functions.

\subsection{\label{ETS_FEDVR}Exterior time scaling with FEDVR functions}

Let $\ell_{\max}$ be the maximum value of $\ell$ in the expansion of $\Psi(\bm{r},t)$ [Eq.~\eqref{Expansion0}]. The radial function for each $\ell$ is then expanded in terms of the FEDVR functions defined in Sec.~\ref{FEDVR function-setup}:
\begin{subequations}
\begin{eqnarray}
\psi_{\ell}(\xi,t)
&=&
\sum_{\hat{\kappa}}
a_{\hat{\kappa}\ell}(t)
\chi_{\hat{\kappa}}(\xi)
+
\sqrt{\frac{2}{1+R(t)}}
a_{\ell}(t)
\chi_{\rm b}(\xi),
\nonumber
\\
&&
\hspace{30mm}
(0\le\xi\le r_{\Sigma}),
\label{orbital expansion0}
\\
\phi_{\ell}(\xi,t)
&=&
\sum_{\check{\kappa}}
b_{\check{\kappa}\ell}(t)
\chi_{\check{\kappa}}(\xi)
+
b_{\ell}(t)
\chi_{\rm b}(\xi),
\nonumber
\\
&&
\hspace{30mm}
(r_{\Sigma}<\xi\le\xi_{\max}),
\label{orbital expansion1}
\end{eqnarray}
\end{subequations}
where the extra factor, $\sqrt{2/[1+R(t)]}$, attached to $a_{\ell}(t)$ is not just cosmetic but helps construct the working equations in Hermitian matrix form (as discussed in the last paragraph in this section). The continuity condition of the wave function, Eq.~\eqref{continuity0}, now leads to
\begin{eqnarray}
\sqrt{\frac{2}{1+R(t)}}
a_{\ell}(t)
&=&
\frac{1}{\sqrt{R(t)}}b_{\ell}(t).
\label{continuity}
\end{eqnarray}
We should thus derive the equations obeyed by the expansion coefficients, $\big\{a_{\hat{\kappa}\ell}(t)\big\}_{\hat{\kappa}\ell}$, $\big\{b_{\check{\kappa}\ell}(t)\big\}_{\check{\kappa}\ell}$, and $\big\{a_{\ell}(t)\big\}_{\ell}$ \big(or alternatively $\big\{a_{\hat{\kappa}\ell}(t)\big\}_{\hat{\kappa}\ell}$, $\big\{b_{\check{\kappa}\ell}(t)\big\}_{\check{\kappa}\ell}$, and $\big\{b_{\ell}(t)\big\}_{\ell}$; in this latter case, the factor, $\sqrt{2/[1+R(t)]}$, attached to $a_{\ell}(t)$ in Eq.~\eqref{orbital expansion0} should be replaced by $1/\sqrt{R(t)}$\big). For simplicity, the rest of this subsection is devoted to the derivation only in the length gauge. The Appendix~\ref{Velocity gauge} lists the instructions for formulating the working equations in the velocity gauge.

It is straightforward to derive the equations for evaluating the time derivative of $a_{\hat{\kappa}\ell}(t)$ and $b_{\hat{\kappa}\ell}(t)$. Substituting Eq.~\eqref{orbital expansion0} into Eq.~\eqref{TDSE_global_TSCSP_in}, and multiplying both sides by $\chi_{\hat{\kappa}}(\xi)$, we integrate them with respect to $\xi$ over $[0,r_{\Sigma}]$ and obtain
\begin{widetext}
\begin{eqnarray}
&&
i\frac{d}{dt}a_{\hat{\kappa}\ell}(t)
=
\sum_{\hat{\kappa}'\ell'}
\bigg\{
\bigg[
\frac{1}{2}
\int_0^{\xi_{\max}}
\frac{d\chi_{\hat{\kappa}}(\xi)}{d\xi}
\frac{d\chi_{\hat{\kappa}'}(\xi)}{d\xi}
d\xi
+
\delta_{\hat{\kappa}\hat{\kappa}'}
V_{\ell}(\xi_{\hat{\kappa}})
\bigg]
\delta_{\ell\ell'}
+
\delta_{\hat{\kappa}\hat{\kappa}'}
W^{\rm L}_{\ell\ell'}(\xi_{\hat{\kappa}},t)
\bigg\}
a_{\hat{\kappa}'\ell'}(t)
\nonumber \\
&&
\hspace{20mm}
+
\frac{a_{\ell}(t)}{\sqrt{2[1+R(t)]}}
\int_0^{\xi_{\max}}
\frac{d\chi_{\hat{\kappa}}(\xi)}{d\xi}
\frac{d \chi_{\rm b}(\xi)}{d \xi}
d\xi.
\label{TSCSP_in} 
\end{eqnarray}
Similarly, substituting Eq.~\eqref{orbital expansion1} into Eq.~\eqref{TDSE_global_TSCSP_out}, and multiplying both sides by $\chi_{\check{\kappa}}(\xi)$, the integration of them with respect to $\xi$ over $(r_{\Sigma},\xi_{\max}]$ results in
\begin{eqnarray}
&&
i\frac{d}{dt}b_{\check{\kappa}\ell}(t)
=
\sum_{\check{\kappa}'\ell'}
\Bigg\{
\delta_{\ell\ell'}
\bigg[
\frac{1}{2[R(t)]^2}
\int_0^{\xi_{\max}}
\frac{d\chi_{\check{\kappa}}(\xi)}{d\xi}
\frac{d\chi_{\check{\kappa}'}(\xi)}{d\xi}
d\xi
+
\delta_{\check{\kappa}\check{\kappa}'}
\bigg(
V_{\ell}\big[r_{\Sigma}+R(t)(\xi_{\check{\kappa}}-r_{\Sigma})\big]
+
\frac{R(t)\ddot{R}(t)}{2}(\xi_{\check{\kappa}}-r_{\Sigma})^2
\bigg)
\bigg]
\nonumber \\
&&
\hspace{20mm}
+
\delta_{\hat{\kappa}\hat{\kappa}'}
W^{\rm L}_{\ell\ell'}[r_{\Sigma}+R(t)(\xi_{\check{\kappa}}-r_{\Sigma}),t]
\Bigg\}b_{\check{\kappa}'\ell}(t)
\frac{a_{\ell}(t)}{\sqrt{2[R(t)]^3[1+R(t)]}}
\int_0^{\xi_{\max}}
\frac{d\chi_{\check{\kappa}}(\xi)}{d\xi}
\frac{d\chi_{\rm b}(\xi)}{d \xi}
d\xi,
\label{TSCSP_out} 
\end{eqnarray}
where Eq.~\eqref{continuity} was used in obtaining the last term of Eq.~\eqref{TSCSP_out}. Note that, after integrating by parts in deriving Eqs.~\eqref{TSCSP_in} and \eqref{TSCSP_out}, every surface term vanishes due to $\chi_{\hat{\kappa}}(0)=\chi_{\hat{\kappa}}(r_{\Sigma})=\chi_{\check{\kappa}}(r_{\Sigma})=\chi_{\check{\kappa}}(\xi_{\max})=0$. Then the integrals, $\int_0^{r_{\Sigma}}$ and $\int_{r_{\Sigma}+0}^{\xi_{\max}}$, respectively, in Eqs.~\eqref{TSCSP_in} and \eqref{TSCSP_out}, are all replaced by $\int_0^{\xi_{\max}}$ because the FEDVR functions, $\chi_{\hat{\kappa}}(\xi)$ and $\chi_{\check{\kappa}}(\xi)$, are, albeit non-vanishing only below and above $r_{\Sigma}$, respectively, defined over the whole interval, $[0,\xi_{\max}]$. 

The derivation of the rest of the equations follows a procedure similar to the above, but requires a little more effort. After the substitution of Eq.~\eqref{orbital expansion0} into Eq.~\eqref{TDSE_global_TSCSP_in}, now multiplying both sides by $2\chi_{\rm b}(\xi)$, we integrate them with respect to $\xi$ over $[0,r_{\Sigma}]$ to obtain
\begin{subequations}
\begin{eqnarray}
&&
i\frac{d}{dt}
\left[
\sqrt{\frac{2}{1+R(t)}}
a_{\ell}(t)
\right]
=
\sum_{\ell'}
\bigg\{
\bigg[
\frac{1}{2}
\int_0^{\xi_{\max}}
\left(
\frac{d\chi_{\rm b}(\xi)}{d\xi}
\right)^2
d\xi
+
V_{\ell}(r_{\Sigma})
\bigg]
\delta_{\ell\ell'}
+
W^{\rm L}_{\ell\ell'}(r_{\Sigma},t)
\bigg\}
\sqrt{\frac{2}{1+R(t)}}
a_{\ell'}(t)
\nonumber \\
&&
\hspace{35mm}
+
\sum_{\hat{\kappa}}
a_{\hat{\kappa}\ell}(t)\int_0^{\xi_{\max}}
\frac{d\chi_{\rm b}(\xi)}{d\xi}
\frac{d\chi_{\hat{\kappa}}(\xi)}{d\xi}
d\xi
-
\chi_{\rm b}(r_{\Sigma})
\frac{\partial}{\partial\xi}
\psi_{\ell}(\xi,t)
\Big|_{\xi=r_{\Sigma}}.
\label{TSCSP_in_b}
\end{eqnarray}
Similarly, substituting Eq.~\eqref{orbital expansion1} into Eq.~\eqref{TDSE_global_TSCSP_out}, multiplying both sides by $2\chi_{\rm b}(\xi)$, and integrating them with respect to $\xi$ over $(r_{\Sigma},\xi_{\max}]$, we arrive at
\begin{eqnarray}
&&
i\frac{d}{dt}b_{\ell}(t)
=
\sum_{\ell'}
\Bigg\{
\bigg[
\frac{1}{2[R(t)]^2}
\int_0^{\xi_{\max}}
\left(
\frac{d\chi_{\rm b}(\xi)}{d\xi}
\right)^2
d\xi
+
V_{\ell}(r_{\Sigma})
\bigg]
\delta_{\ell\ell'}
+
W^{\rm L}_{\ell\ell'}(r_{\Sigma},t)
\Bigg\}
b_{\ell'}(t)
\nonumber \\
&&
\hspace{14mm}
+
\sum_{\check{\kappa}}
\frac{b_{\check{\kappa}\ell}(t)}{[R(t)]^2}
\int_0^{\xi_{\max}}
\frac{d\chi_{\rm b}(\xi)}{d\xi}
\frac{d\chi_{\check{\kappa}}(\xi)}{d\xi}
d\xi
+
\frac{1}{[R(t)]^{2}}
\chi_{\rm b}(r_{\Sigma})
\frac{\partial}{\partial\xi}
\phi_{\ell}(\xi,t)
\Big|_{\xi=r_{\Sigma}+0}.
\label{TSCSP_out_b}
\end{eqnarray}
\end{subequations}
In obtaining Eqs.~\eqref{TSCSP_in_b} and \eqref{TSCSP_out_b}, we used the relation,
\begin{eqnarray}
2
\int_0^{r_{\Sigma}}
\left(
\frac{d\chi_{\rm b}(\xi)}{d\xi}
\right)^2
d\xi
=
\int_0^{\xi_{\max}}
\left(
\frac{d\chi_{\rm b}(\xi)}{d\xi}
\right)^2
d\xi
=
2
\int_{r_{\Sigma}}^{\xi_{\max}}
\left(
\frac{d\chi_{\rm b}(\xi)}{d\xi}
\right)^2
d\xi,
\end{eqnarray}
\end{widetext}
as well as the DVR quadrature rule for the bridge function: For an arbitrary integrable function $f(\xi)$ defined in $[0,\xi_{\max}]$,
\begin{subequations}
\label{DVR quadrature rule}
\begin{eqnarray}
&&2
\int_0^{r_{\Sigma}}
\chi_{\rm b}(\xi)
f(\xi)
\chi_{\kappa}(\xi)
d\xi
\approx
\int_0^{\xi_{\max}}
\chi_{\rm b}(\xi)
f(\xi)
\chi_{\kappa}(\xi)
d\xi
\nonumber
\\
&\approx&
2
\int_{r_{\Sigma}}^{\xi_{\max}}
\chi_{\rm b}(\xi)
f(\xi)
\chi_{\kappa}(\xi)
d\xi,
\label{DVR quadrature rule a}
\end{eqnarray}
and 
\begin{eqnarray}
\hspace{-3mm}
\int_0^{\xi_{\max}}
\chi_{\rm b}(\xi)
f(\xi)
\chi_{\kappa}(\xi)
d\xi
\approx
\delta_{\kappa,(N_{\rm dvr}-1)\hat{N}_{\rm fe}}
f(r_{\Sigma}),
\label{DVR quadrature rule b}
\end{eqnarray}
\end{subequations}
where all the almost-equal signs in Eq.~\eqref{DVR quadrature rule} become exact-equal signs if $f(\xi)$ is linear in $\xi$ in $\big(\xi_{\hat{N}_{\rm fe}-1},\xi_{\hat{N}_{\rm fe}}\big]\cup\big(\xi_{\hat{N}_{\rm fe}},\xi_{\hat{N}_{\rm fe}+1}\big]$ (see, e.g., Ref.~\cite{Light1985}). Differently from Eqs.~\eqref{TSCSP_in} and \eqref{TSCSP_out}, Eqs.~\eqref{TSCSP_in_b} and \eqref{TSCSP_out_b} contain the surface terms at $r_{\Sigma}$ because of $\chi_{\rm b}(r_{\Sigma})\ne0$. In dealing with the surface terms, the Bloch operator, $\mathcal{L}=\frac{1}{2}\delta(\xi-r_{\Sigma})\frac{\partial}{\partial \xi}$, is a useful device as in the formulation of $R$-matrix-related theories (see, e.g., Refs.~\cite{Descouvemont2010,Burke2011}). In the interest of keeping the derivation mathematically clear, however, we do not employ the Bloch operator the delta function of which could cause an ambiguity about whether $\Sigma$ belongs to its inside or outside or, perhaps, both sides or neither side. Getting back to the derivation, canceling out the surface terms in Eqs.~\eqref{TSCSP_in_b} and \eqref{TSCSP_out_b} using Eq.~\eqref{diff_continuity}, we finally obtain
\begin{eqnarray}
&&
i\frac{d}{dt}a_{\ell}(t)
=
\sum_{\ell'}
\bigg\{
\bigg[
\frac{1}{2R(t)}
\int_{0}^{\xi_{\max}}
\left(
\frac{d\chi_{\rm b}(\xi)}{d\xi}
\right)^2
d\xi
\nonumber \\
&&
\hspace{5mm}
+
V_{\ell}(r_{\Sigma})
\bigg]
\delta_{\ell\ell'}
+
W^{\rm L}_{\ell\ell'}(r_{\Sigma},t)
\bigg\}
a_{\ell'}(t)
\nonumber
\\
&&
\hspace{5mm}
+
\sum_{\hat{\kappa}}
\frac{a_{\hat{\kappa}\ell}(t)}{\sqrt{2[1+R(t)]}}\int_0^{\xi_{\max}}
\frac{d\chi_{\rm b}(\xi)}{d\xi}
\frac{d\chi_{\hat{\kappa}}(\xi)}{d\xi}
d\xi
\nonumber \\
&&
\hspace{5mm}
+
\sum_{\check{\kappa}}
\frac{b_{\check{\kappa}\ell}(t)}{\sqrt{2[R(t)]^3[1+R(t)]}}
\int_{0}^{\xi_{\max}}
\frac{d\chi_{\rm b}(\xi)}{d\xi}
\frac{d \chi_{\check{\kappa}}(\xi)}{d\xi}
d\xi,
\nonumber \\
\label{TSCSP_in_out_b} 
\end{eqnarray}
where Eq.~\eqref{continuity} was used to express $b_{\ell}(t)$ in terms of $a_{\ell}(t)$.

Assembling the expansion coefficients, $\big\{a_{\hat{\kappa}\ell}(t)\big\}_{\hat{\kappa}\ell}$, $\big\{b_{\check{\kappa}\ell}(t)\big\}_{\check{\kappa}\ell}$, and $\big\{a_{\ell}(t)\big\}_{\ell}$, into a vector, $\bm{\Psi}(t)$, the set of working equations, Eqs.~\eqref{TSCSP_in}, \eqref{TSCSP_out}, and \eqref{TSCSP_in_out_b}, is expressed in matrix form, $i\dot{\bm{\Psi}}(t)={\bf H}(t)\bm{\Psi}(t)$, where ${\bf H}(t)$ is real and symmetric in the length gauge (Hermitian in the velocity gauge; see Appendix~\ref{Velocity gauge}). Hence the short-time iterative Lanczos method~\cite{Park1986,Kuleff2005} is a very efficient algorithm for the time propagation. Note that, in the numerical implementation, several elements of ${\bf H}(t)$ attached by $R(t)$ need update at each time step. The CPU time for this extra operation, which is absent in the usual non-time scaled method, is, however, not so large and causes no major problem. Section~\ref{HHG} shows the efficiency of ETS compared with the non-time scaled method based on practical test calculations. 

Finally note the normalization condition of the wave function:
\begin{eqnarray}
&&
\int
|\Psi({\bm r},t)|^2
d{\bm r}
\nonumber \\
&&
\hspace{5mm}
=
\sum_{\ell}
\left[
\int_0^{r_{\Sigma}}
|\psi_{\ell}(\xi,t)|^2
d\xi
+
\int_{r_{\Sigma}+0}^{\xi_{\max}}
|\phi_{\ell}(\xi,t)|^2
d\xi
\right]
\nonumber
\\
\nonumber \\
&&
\hspace{5mm}
=
\sum_{\ell}
\left[
\sum_{\hat{\kappa}}|a_{\hat{\kappa}\ell}(t)|^2
+
|a_{\ell}(t)|^2
+
\sum_{\check{\kappa}}|b_{\check{\kappa}\ell}(t)|^2
\right]
\nonumber \\
&&
\hspace{5mm}
=
\|\bm{\Psi}(t)\|^2
=1,
\label{Normalization}
\end{eqnarray}
where Eq.~\eqref{continuity} was used in obtaining the third line. The extra factor, $\sqrt{2/[1+R(t)]}$, attached to $a_{\ell}(t)$ in Eq.~\eqref{orbital expansion0} serves to provide the conclusion of Eq.~\eqref{Normalization}: $\int|\Psi({\bm r},t)|^2d{\bm r}=\|\bm{\Psi}(t)\|^2=1$. In the formulation without this factor, we arrive at another normalization condition, $\int|\Psi({\bm r},t)|^2d{\bm r}=\sum_{\ell}\big[\sum_{\hat{\kappa}}|a_{\hat{\kappa}\ell}(t)|^2+[1+R(t)]|a_{\ell}(t)|^2/2+\sum_{\check{\kappa}}|b_{\check{\kappa}\ell}(t)|^2\big]=1\ne \|\bm{\Psi}(t)\|^2$, which indicates the decrease of $\|\bm{\Psi}(t)\|^2$ as $R(t)$ grows. To fulfill the normalization condition, a complex term, $-i\dot{R}(t)/\{2[1+R(t)]\}a_{\ell}(t)$, shows up in an equation corresponding to Eq.~\eqref{TSCSP_in_out_b} on its right hand side. Consequently, the working equations turn out to be non-Hermitian in matrix form if the factor $\sqrt{2/[1+R(t)]}$ is not included. The time propagation can still be implemented by the Arnoldi algorithm~\cite{Arnoldi1951,Kuleff2005} or the Runge-Kutta method~\cite{Press2007}, for instance, but not by the simple Lanczos algorithm. The factor, $\sqrt{2/[1+R(t)]}$, is hence better attached to $a_{\ell}(t)$ in Eq.~\eqref{orbital expansion0} to prevent such unnecessary complications.

\section{\label{Stiffness-free Lanczos algorithm}Stiffness-free FEDVR-based Lanczos algorithm}

As seen from the formulation in Sec.~\ref{Formulation} and the explicit form of the working equations of Sec.~\ref{Practical}, ETS is a generalization of GTS aiming at the reduction of stiffness; non-time scaling in the inner region avoids shrinking of most bound states, and hence, differently from GTS, allows the use of less dense basis functions around the origin. In the treatment of strong-field ionization at mid-IR wavelengths, corresponding to the tunneling regime, the equations of motion in ETS inevitably still become stiff as many angular momentum states are involved. To resolve this remaining problem, this section provides a detailed analysis to identify the origin of the stiffness, and proposes a procedure for its removal. Note that this section holds some independence from the other sections. The discussion in the following is not only applicable to ETS implementation, but also to a wider class of equations of motion appearing in atomic and molecular physics.

\subsection{\label{Error assessment}Error and stiffness analysis on Lanczos algorithm}

\begin{figure*}
\begin{center}
\begin{tabular}{c}
\resizebox{175mm}{!}{\includegraphics{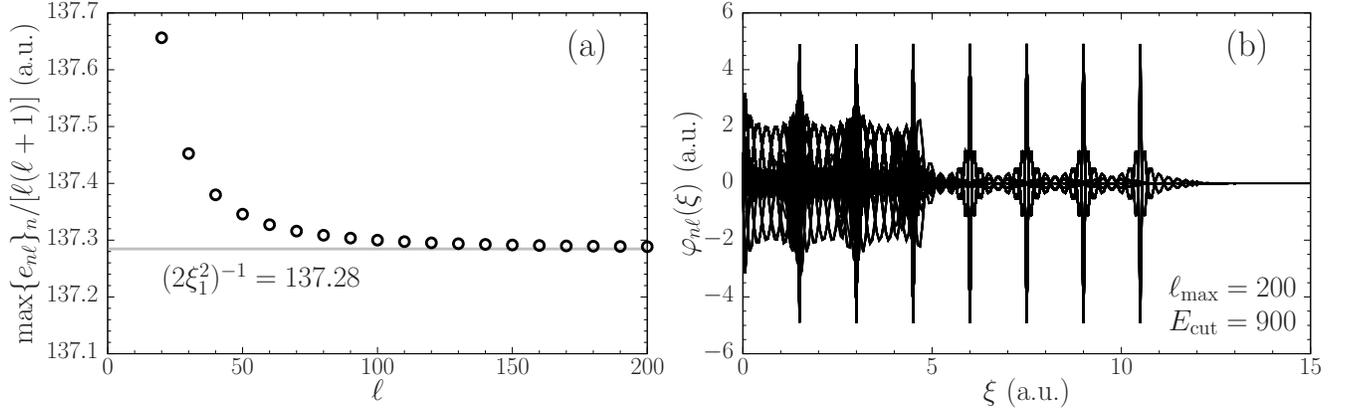}}
\end{tabular}
\caption{
\label{stiff}
(a) Plot of $\max\{e_{n\ell}\}_{n}/[\ell(\ell+1)]$ as a function of $\ell$ (in this case, $\max\{e_{n\ell}\}_{n}=e_{89,\ell}$). Every ten data is shown by open circles. The solid gray line indicates $(2\xi_1^2)^{-1}=137.28$. (b) Eigenvectors of ${\bf h}_{\ell}$, satisfying $e_{n\ell}>E_{\rm cut}=900$ ($\ell=0,\cdots,200(=\ell_{\max})$). The plot is made in terms of the FEDVR expansion, $\varphi_{n\ell}(\xi)\equiv \sum_{\hat{\kappa}=1}^{\tilde{N}}({\bf u}_{\ell})_{n\hat{\kappa}}\chi_{\hat{\kappa}}(\xi)$. Both (a) and (b) are obtained under the same numerical condition: $\tilde{N}=89$ ($N_{\rm dvr}=10$ and $\tilde{N}_{\rm fe}=10$) and $\Delta \xi=1.5$.
}
\end{center}
\end{figure*}

\begin{figure*}
\begin{center}
\begin{tabular}{c}
\resizebox{160mm}{!}{\includegraphics{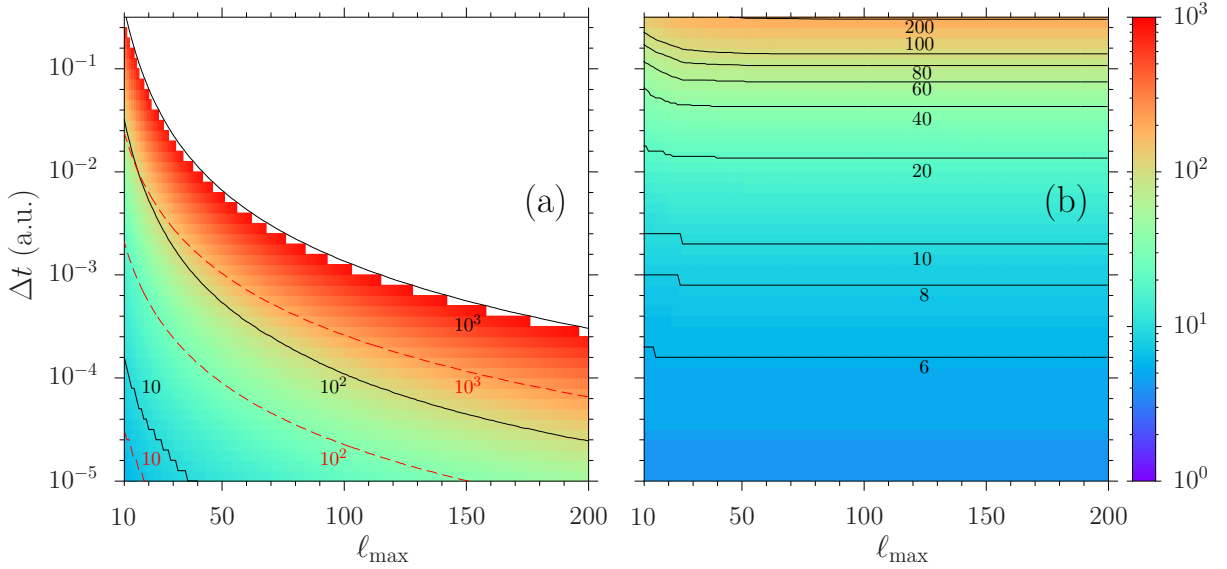}}
\end{tabular}
\caption{
\label{kry01}
(Color online) (a) Contour plot of $K_{\max}$ as a function of $\ell_{\max}$ and $\Delta t$. The color code and solid black lines represent the contour based on Eq.~\eqref{error 0} with $\epsilon=10^{-15}$. The blank area indicates $K_{\max}>1000$. This estimation is obtained for field-free Hamiltonian, ${\bf H}(0)$ under the numerical condition: $r_{\Sigma}=\xi_{\max}=60$, $N_{\rm dvr}=10$, $\hat{N}_{\rm fe}=40$, $\check{N}_{\rm fe}=0$, and $\Delta\xi=1.5$. The dashed (red) lines show the contour based on Eq.~\eqref{error 5}. (b) The same as (a) except that the stiffness is removed by setting $\tilde{N}_{\rm fe}=10$ and $E_{\rm cut}=900$. See the main text for details.
}
\end{center}
\end{figure*}

This subsection focuses on stiffness (see, e.g., Ref.~\cite{Press2007}), i.e., the degree of difficulty in a numerical treatment of the working equations, to see how it poses problems, and to identify its origin for seeking the resolution. Toward this end, the discussion commences with an error analysis on the short-time iterative Lanczos algorithm. For a given vector, $\bm{\Psi}(t)$, the evolution in a small time step $\Delta t$ is given by
\begin{eqnarray}
\bm{\Psi}(t+\Delta t)
=
\sum_{k=0}^{K-1}
\frac{1}{k!}
\big[-i{\bf H}(t)\Delta t\big]^k
\bm{\Psi}(t),
\label{Lanczos}
\end{eqnarray}
with error 
\begin{eqnarray}
&&\|{\bm \Psi}_{\rm exact}(t+\Delta t)-{\bm \Psi}(t+\Delta t)\|^2
\nonumber
\\
&&
\hspace{10mm}
\approx
\left|
\frac{\beta_1(t)\beta_2(t)\cdots \beta_{K-1}(t)}{(K-1)!}
(\Delta t )^{K-1}
\right|^2
<\epsilon,
\label{error 0}
\end{eqnarray}
where $\{\beta_k(t)\}_{k=1}^{K-1}$ is a set of subdiagonal elements in a reduced form of ${\bf H}(t)$ on the $K$-dimensional Krylov subspace~\cite{Park1986}, and the last inequality requires the error to be less than, say, $\epsilon\equiv10^{-15}$. Setting $\Delta t$ much smaller than the characteristic time scale of external fields, Eq.~\eqref{error 0} serves as a criterion at every time step to determine the smallest integer for $K$. If the time propagation starts with the ground state, the product $\beta_1(t)\beta_2(t)\cdots \beta_{K-1}(t)$ is zero at $t=0$, but in due course reaches its maximum, when the Krylov subspace acquires the largest dimension, $K_{\max}$; the computation faces difficulty if $K_{\max}$ is too large. Let $B$ denote the maximum of $\sqrt[K-1]{\beta_1(t)\beta_2(t)\cdots \beta_{K-1}(t)}$ during time evolution. Note that, although field-free atomic states, the energy eigenvalues of which are well above the energy range of physical interest, could participate in construction of the Krylov subspace, the population of such high-energy states are virtual excitations which happen more likely for larger $\Delta t$. Supposing the worst case such that every energy eigenstate is accessed in the virtual excitations, let the analysis in the following be based upon an assumption, $B=\max\{E_{n\ell}\}_{n\ell}$, i.e., $B$ reaches the maximum eigenvalue of ${\bf H}(0)$~
\footnote
{
For the time-independent Hamiltonian, i.e., if there is no light field so that ${\bf H}(t)={\bf H}(0)$, it can be shown that $B< \sqrt[K-1]{2} \max\{E_{n\ell}\}_{n\ell} \exp\big[{\max\{E_{n\ell}\}_{n\ell}/(K-1)}\big]$. Additionally, if ${\bf H}(0)$ is positive definite, i.e., if ${\bf H}(0)$ holds no bound state, it can also be shown that $B<  2^{(K-2)/(K-1)} \max\{E_{n\ell}\}_{n\ell}$. The latter is not the case in our discussion, but may give better estimation to our error and stiffness assessment, because the stiffness mainly arises from very large angular momentum states supporting almost no bound state. See Refs.~\cite{Gallopoulos1992,Saad1992} for mathematical details.
}. In this assumption, the working equations become stiffer as $\max\{E_{n\ell}\}_{n\ell}$ increases. 

Meanwhile, looking at the working equations~\eqref{TSCSP_in}, \eqref{TSCSP_out}, and \eqref{TSCSP_in_out_b}, we notice that $V_{\ell}(\xi_{\hat{\kappa}})$ in Eq.~\eqref{TSCSP_in} takes very large values in the vicinity of the nucleus [see Eq.~\eqref{atomic potential}] and is the decisive factor of $\max\{E_{n\ell}\}_{n\ell}$; hence, Eq.~\eqref{TSCSP_in} is responsible for the stiffness. To discuss more quantitatively, we define a set of field-free Hamiltonian matrices around the nucleus,
\begin{eqnarray}
&&\big({\bf h}_{\ell}\big)_{\hat{\kappa}\hat{\kappa}'}
\equiv
\frac{1}{2}
\int_0^{\xi_{\max}}
\frac{d\chi_{\hat{\kappa}}(\xi)}{d\xi}\frac{d\chi_{\hat{\kappa}'}(\xi)}{d\xi}
d \xi
+
\delta_{\hat{\kappa}\hat{\kappa}'}
V_{\ell}(\xi_{\hat{\kappa}}),
\nonumber
\\
&&\hspace{15mm}
\big(
\hat{\kappa},\hat{\kappa}'=1,\cdots,\tilde{N},
\hspace{2mm}
\ell=0,\cdots,\ell_{\max}
\big),
\label{h_kl}
\end{eqnarray}
where $\tilde{N}=(N_{\rm dvr}-1)\tilde{N}_{\rm fe}-1$ with $\tilde{N}_{\rm fe}\le \hat{N}_{\rm fe}$, and note the boundary condition, $\chi_{\hat{\kappa}}(\xi(\ge\xi_{\tilde{N}}))=0$ for $\hat{\kappa}=1,\cdots,{\tilde{N}}$. Diagonalizing these small matrices and obtaining ${\bf u}_{\ell}^{\rm T}{\bf h}_{\ell}{\bf u}_{\ell}={\rm diag} \big(e_{1\ell},e_{2\ell},\cdots,e_{\tilde{N}\ell}\big)$, the eigenvalues for a first few integers of $n$ represent the bound-state energies, $e_{n\ell}\approx- [2(n+\ell)^{2}]^{-1}$. The rest of the eigenvalues are positive and could, in particular for large $\ell$, be too large to be of importance in the physical process of interest. We then suppose
\begin{eqnarray}
\max\{E_{n\ell}\}_{n\ell}
\approx 
\max\{e_{n\ell}\}_{n\ell}
\approx 
\frac{\ell_{\max}(\ell_{\max}+1)}{2\xi_1^2},
\label{error 4}
\end{eqnarray}
i.e., $\max\{e_{n\ell}\}_{n\ell}$ is not sensitive to $\tilde{N}$ and almost determined by the centrifugal part of $V_{\ell_{\max}}(\xi_1)$ in Eq.~\eqref{atomic potential}. This conjecture is verified by numerical examples: Setting $\tilde{N}=89$ ($N_{\rm dvr}=10$ and $\tilde{N}_{\rm fe}=10$) and $\Delta \xi=1.5$, the greatest eigenvalue for each $\ell$, $\max\{e_{n\ell}\}_{n}=e_{89,\ell}$, is, after divided by $\ell(\ell+1)$, plotted in Fig.~\ref{stiff} (a). This kind of plot is, for fixed $N_{\rm dvr}=10$ and $\Delta \xi=1.5$, insensitive to the change of $\tilde{N}_{\rm fe}$ (not shown in the figure), and, in every case, approaching $(2\xi_1)^{-1}=137.28$ at large $\ell$. Hence, using Eq.~\eqref{error 4} in Eq.~\eqref{error 0} with assuming $B=\max\{E_{n\ell}\}_{n\ell}$ gives
\begin{eqnarray}
&&
\frac{K_{\max}-1}{\Delta t}
>
e
\cdot
[2\pi(K_{\max}-1)\epsilon]^{-1/[2(K_{\max}-1)]}
\nonumber
\\
&&
\hspace{20mm}
\times
\frac{\ell_{\max}(\ell_{\max}+1)}{2\xi_1^2},
\label{error 5}
\end{eqnarray}
where $(K_{\max}-1)!>\sqrt{2\pi (K_{\max}-1)}\cdot[(K_{\max}-1)/e]^{K_{\max}-1}$ is used (Stirling's formula; see, e.g., Ref.~\cite{Abramowitz1972}). Noting that the numerical cost is proportional to $(K_{\max}-1)/\Delta t$, and also $[2\pi(K_{\max}-1)\epsilon]^{-1/[2(K_{\max}-1)]}\to 1+0$ as $K_{\max}\to\infty$, Eq.~\eqref{error 5} indicates the growing numerical efficiency by setting $\Delta t$ as large as possible (as long as it is still much smaller than the characteristic time scale of external fields). Equation~\eqref{error 5} at the same time manifests the overwhelming stiffness for large $\ell_{\max}$ and/or small $\xi_1$. Although the assumption $B=\max\{E_{n\ell}\}_{n\ell}$ is so na\"ive that Eq.~\eqref{error 5} may overestimate $K_{\max}$, a more rigorous theoretical analysis is beyond the scope of this paper. Let us verify Eq.~\eqref{error 5} numerically instead; Fig.~\ref{kry01}(a) shows $K_{\max}$ as a function of $\ell_{\max}$ and $\Delta t$. The color code and solid black lines represent the contour based on Eq.~\eqref{error 0} with $\epsilon=10^{-15}$; $K_{\max}$ is calculated for each pair of $\ell$ and $\Delta t$ as $\max\{K_i\}_{i=1}^{100}$, where $K_i$ is the smallest integer satisfying Eq.~\eqref{error 0} in obtaining $\bm{\Psi}_i(\Delta t)$ from $\bm{\Psi}_i(0)$ which is every time ($i=1,\cdots,100$) constructed by random-number elements and normalized. The calculation is based on a field-free non-time-scaled Hamiltonian matrix ${\bf H}(0)$ with the following parameters: $r_{\Sigma}=\xi_{\max}=60$, $N_{\rm dvr}=10$, $\hat{N}_{\rm fe}=40$, $\check{N}_{\rm fe}=0$, and $\Delta\xi=1.5$. The same calculation with a Hamiltonian matrix including light-atom interaction term (with fixed light intensity $I=10^{14}$ W/cm$^2$) in the length gauge causes no visible change. The dashed (red) lines in Fig.~\ref{kry01}(a), representing the contour based on Eq.~\eqref{error 5}, show obvious overestimation but capture the gross feature of the landscape, verifying Eq.~\eqref{error 5} despite its simplicity.

Equation~\eqref{error 5} provides explicitly the following insights: Achieving high accuracy in the description of electronic structure and tunneling dynamics requires many FEDVR functions in the fixed interval, $[0,\xi_{\max}]$, as well as many angular momentum states. With improved accuracy, however, $\ell_{\max}(\ell_{\max}+1)/(2\xi_1^2)$ rapidly increases, requiring larger $K_{\max}$, which leads to a growing stiffness in the working equations. Note that the stiffness problem is not peculiar in atomic systems with the FEDVR-based Lanczos algorithm, but rather inherently appears in a variety of differential equations in numerical treatments.

\subsection{\label{Stiffness}Stiffness removal}

\begin{figure*}
\begin{center}
\begin{tabular}{c}
\resizebox{120mm}{!}{\includegraphics{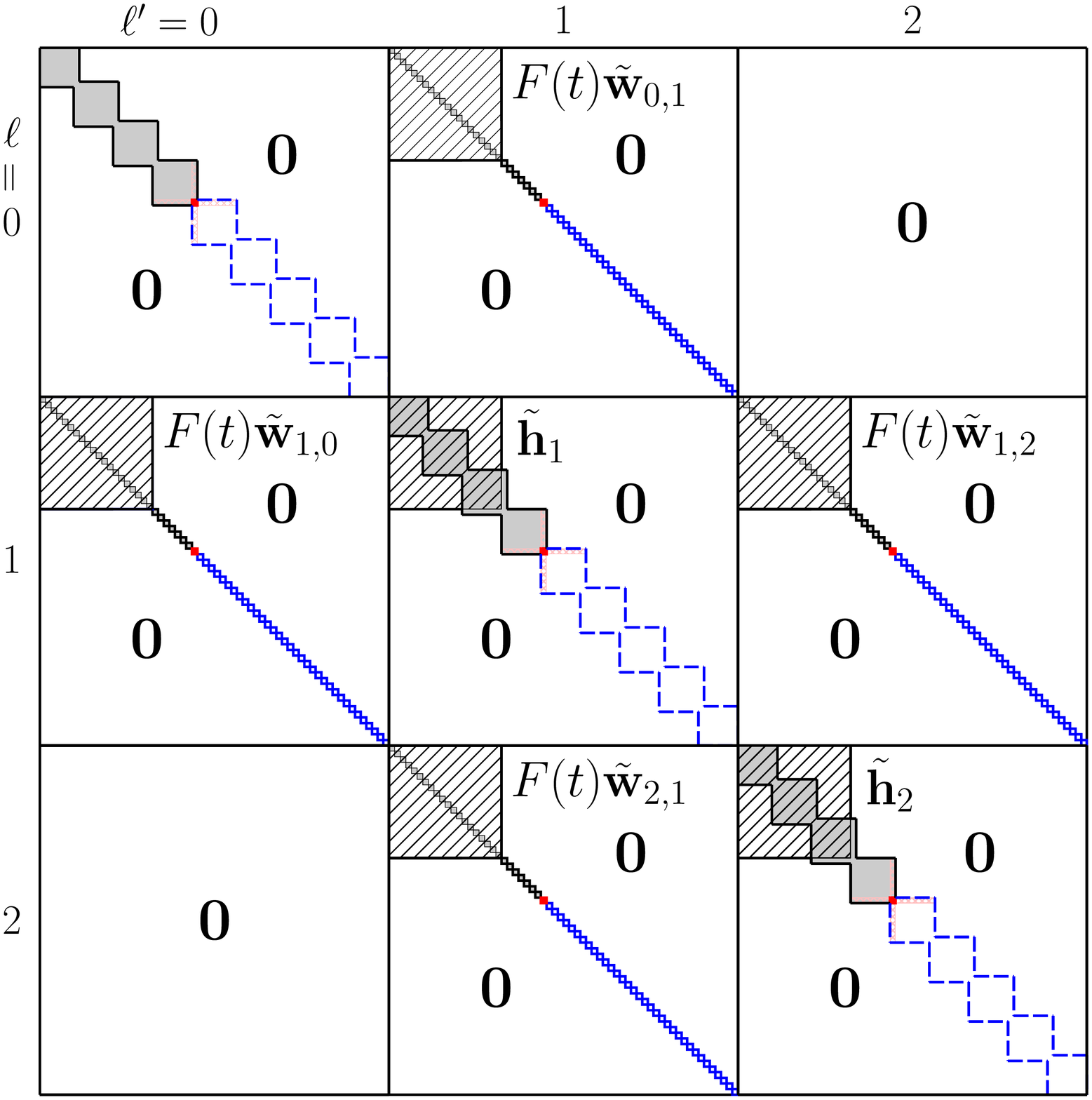}}
\end{tabular}
\caption{
\label{fedvr_matrix}
(Color online) Structure of the stiffness-free matrix, $\tilde{\bf H}(t)$. This is an illustrative example in the length gauge for the first three angular momentum states ($\ell=0,1,$ and $2$) in a situation of ${\bf h}_{0}=\tilde{\bf h}_{0}$ and ${\bf h}_{\ell}\ne \tilde{\bf h}_{\ell}$ ($\ell\ge 1$). Before the stiffness removal, each angular momentum diagonal block ($\ell=\ell'$) has block-diagonal structure with respect to $\xi$, and each angular momentum subdiagonal block ($\ell=|\ell'\pm 1|$) is just diagonal (because of the length gauge). After the stiffness removal, ${\bf h}_{\ell}$ is replaced by $\tilde{\bf h}_{\ell}$ for $\ell\ge1$ [see Eq.~\eqref{h_sf}], and every light-atom interaction block $F(t){\bf w}_{\ell\ell'}$ is replaced by $F(t)\tilde{\bf w}_{\ell\ell'}$ [see Eq.~\eqref{w_sf}]. Consequently, the top-left corners of angular momentum diagonal and subdiagonal blocks become full. This figure is consistent with Fig.~\ref{fedvr_basis} in color code; the shaded, non-shaded (blue), and filled (red) parts, respectively, indicate the matrix elements formed by $\big\{\chi_{\hat{\kappa}}(\xi)\big\}_{\hat{\kappa}}$, $\big\{\chi_{\rm b}(\xi)\big\}$, and $\big\{\chi_{\check{\kappa}}(\xi)\big\}_{\check{\kappa}}$.
}
\end{center}
\end{figure*}

As shown in Sec.~\ref{Error assessment}, the stiffness is mainly caused by high-angular momentum states, and due to the large value of the centrifugal potential around the vicinity of the nucleus. This conclusion suggests a clear strategy to remove the stiffness; setting a cutoff energy $E_{\rm cut}$ well above the range of physical interest, the eigenvectors of ${\bf h}_{\ell}$ should be excluded from the calculation if $e_{n\ell}>E_{\rm cut}$. Figure~\ref{stiff} (b) depicts eigenvectors of ${\bf h}_{\ell}$, the eigenvalues of which are above the cutoff, $e_{n\ell}>E_{\rm cut}=900$. Such high-energy states localize only around the nucleus, hence they can be safely excluded as follows. Defining a set of $\tilde{N}\times \tilde{N}$ matrices by
\begin{eqnarray}
&&\big(\tilde{\bf u}_{\ell}\big)_{n\hat{\kappa}}
\equiv
\left\{
\begin{array}{lcl}
\big({\bf u}_{\ell}\big)_{n\hat{\kappa}}& &  {\rm for\;} e_{n\ell}\le E_{\rm cut}, \\ \\
0& &   {\rm for\;} e_{n\ell}> E_{\rm cut}, \\
\end{array}
\right.
\nonumber
\\
&&
\hspace{15mm}
\big(
n,\hat{\kappa}=1,\cdots,\tilde{N};
\hspace{2mm}
\ell=0,\cdots,\ell_{\max}
\big),
\end{eqnarray}
let us approximately reconstruct ${\bf h}_{\ell}$ as
\begin{eqnarray}
{\bf h}_{\ell}
&=&
{\bf u}_{\ell} {\bf u}^{\rm T}_{\ell}
{\bf h}_{\ell}
{\bf u}_{\ell} {\bf u}^{\rm T}_{\ell}
\nonumber
\\
&\approx&
\tilde{\bf u}_{\ell} {\bf u}^{\rm T}_{\ell} {\bf h}_{\ell}{\bf u}_{\ell} \tilde{\bf u}^{\rm T}_{\ell}
\equiv
\tilde{\bf h}_{\ell}.
\label{h_sf}
\end{eqnarray}
The light-atom interaction should accordingly be modified around the nucleus; canceling out the TD field factor in the light-atom interaction operator, and defining time-independent matrix by [see Eq.~\eqref{light_attom_interaction}]
\begin{eqnarray}
&&
\big({\bf w}_{\ell\ell'}\big)_{\hat{\kappa}\hat{\kappa}'}
\equiv
\frac{1}{G(t)}
\int_0^{\xi_{\max}}
\chi_{\hat{\kappa}}(\xi)
W_{\ell\ell'}(t)
\chi_{\hat{\kappa}'}(\xi)
d \xi,
\nonumber
\\
&&
\hspace{12mm}
\big(
\hat{\kappa},\hat{\kappa}'=1,\cdots,\tilde{N},
\hspace{2mm}
\ell,\ell'=0,\cdots,\ell_{\max}
\big),
\label{w_kl}
\end{eqnarray}
with $G(t)$ denoting $F(t)$ ($A(t)$) in the length (velocity) gauge, let Eq.~\eqref{w_kl} then be approximated as
\begin{eqnarray}
{\bf w}_{\ell\ell'}
&=&
{\bf u}_{\ell} {\bf u}^{\rm T}_{\ell}
{\bf w}_{\ell\ell'}
{\bf u}_{\ell'} {\bf u}^{\rm T}_{\ell'}
\nonumber
\\
&\approx&
\tilde{\bf u}_{\ell} {\bf u}^{\rm T}_{\ell}{\bf w}_{\ell\ell'}{\bf u}_{\ell'} \tilde{\bf u}^{\rm T}_{\ell'}
\equiv
\tilde{\bf w}_{\ell\ell'}.
\label{w_sf}
\end{eqnarray}
The set of working equations is now approximated as $i\dot{\bm{\Psi}}(t)=\tilde{\bf H}(t)\bm{\Psi}(t)$, where $\tilde{\bf H}(t)$ is constructed using $\tilde{\bf h}_{\ell}$ and $G(t)\tilde{\bf w}_{\ell\ell'}$ (see Fig.~\ref{fedvr_matrix}) and is hence expected to be stiffness free and to lead to a reduction in $K_{\max}$. Look at the contour plot of $K_{\max}$ in Fig.~\ref{kry01}(b), which is computed in the same numerical condition as in Fig.~\ref{kry01}(a) after the application of stiffness removal procedure with $\tilde{N}_{\rm fe}=10$ and $E_{\rm cut}=900$. Figure~\ref{kry01}(b) confirms our expectation.

Note that ${\bf h}_{\ell}$ is structured block diagonal, and ${\bf w}_{\ell}$ is diagonal (block diagonal) in the length (velocity) gauge, whereas both $\tilde{\bf h}_{\ell}$ and $\tilde{\bf w}_{\ell\ell'}$ are full. Hence, the stiffness removal partly destroys the sparseness of ${\bf H}(t)$, but $\tilde{\bf H}(t)$ is still largely sparse and can be efficiently handled in Harwell-Boeing format (see, e.g., Ref.~\cite{Press2007}). In most cases, $\tilde{N}_{\rm fe}(\le\hat{N}_{\rm fe})$ will be set around $10$ to safely remove the stiffness. Setting $\tilde{N}_{\rm fe}$ smaller makes $\tilde{\bf H}(t)$ sparser. However, we always need to check before starting the time propagation that the removed high-energy states do not localize near the right edge of this small interval (around $\xi=15$ in the case of Fig.~\ref{stiff}(b)). If not, it is safe, but if so, $\tilde{N}_{\rm fe}$ and/or $E_{\rm cut}$ must be set larger.

The stiffness removal procedure given above is not easily applicable to GTS because in GTS every element of the Hamiltonian matrix depends on time. Also note that the stiffness removal itself is not a new concept. One can find a discussion for the (without-space-partition) TD B-spline $R$-matrix approach in Ref.~\cite{Guan2008}. Similar procedures have thereafter been proposed by several researchers. Reference~\cite{Hochstuhl2014}, for instance, suggests a procedure for TD many-electron calculations based on the FEDVR functions. The stiffness is removed, however, over the whole spatial simulation volume. The total Hamiltonian matrix becomes completely full, spoiling the benefit of FEDVR functions. One can see another procedure more similar to ours in Ref.~\cite{Hart2014_Ne+}, where the authors investigate the photoionization of Ne$^+$ by means of RMT~\cite{Nikolopoulos2008,Moore2011a,Hart2014}. Based upon the space-partition concept, the radial wave functions are expressed in terms of the $R$-matrix basis functions and grids, respectively, inside and outside a spherical sphere the radius of which is $r_{\Sigma}=15$. The working equations are free from stiffness by setting $E_{\rm cut}=1345$ and composing the $R$-matrix basis functions of field-free eigenstates.

\section{\label{Ionization}Numerical demonstration}

By solving the ETS working equations \eqref{TSCSP_in}, \eqref{TSCSP_out}, and \eqref{TSCSP_in_out_b} with the stiffness-free procedure discussed in Sec.~\ref{Stiffness}, let atomic hydrogen, prepared in the ground state at $t=0$, time evolve under an $N$-cycle mid-IR ($\lambda=3$ $\mu$m $=5.669\times10^{4}$) pulse the envelop of which is defined by
\begin{eqnarray}
f_N(t)
=
\left\{
\begin{array}{lcl}
\displaystyle{
\sin^2\big(\pi t/T\big)
}
& &  (0\le t \le T) \\ \\
\displaystyle{
0
}
& &(T<t) \\
\end{array}
\right.
\label{f(t)}
\end{eqnarray}
where $T=2\pi N/\omega=413.7\times N$. 
In due care of Eq.~\eqref{f(t)}, the time-scaling factor is now defined by
\begin{eqnarray}
R(t)
=
\left\{
\begin{array}{lc}
\displaystyle{
\frac{{R_{\infty}}}{2T}
\left\{
t^2
+
\frac{T^2}{2\pi^2}
\left[
\cos\bigg(\frac{2\pi t}{T}\bigg)
-1
\right]
\right\}
+1
}&
\\
\hspace{39mm}  (0\le t \le T) \\ \\
\displaystyle{
{R_{\infty}}(t-T)
+
\frac{T{R_{\infty}}}{2}
+1
}
\hspace{10mm}  
(T<t) \\
\end{array}
\right.
\label{R(t)_2}
\end{eqnarray}
which gives non-vanishing $\ddot{R}(t)$ only during the presence of the light field: $\ddot{R}(t)=(2{R_{\infty}}/T)f_N(t)$ for $0\le t \le T$, and $\ddot{R}(t)=0$ for $t>T$. The original idea of time scaling appeared in search of suitable adiabatic parameters for describing atom-diatom collisions~\cite{Solovievt1985}. In our case, supposing some adiabatic action of the mid-IR pulse on the electron, the time-scaling factor will be better specified by the carrier envelop function. Note that Refs.~\cite{Sidky2000,Derbov2003,Roudnev2005,Serov2007,Serov2008,Hamido2011,Frapiccini2015} propose another form, $R(t)=[1+({R_{\infty}}(t-t_0))^n]^{1/n}$ $(n=2,3$, or $4)$, for investigating ionization by high-energy photoabsorption (in the extreme ultraviolet region) and fast-electron impact (in the keV region); Ref.~\cite{Hamido2011} also reports the insensitivity of the calculation to the starting time $t_0$. For mid-IR laser pulses, however, this definition is not well suited because $R(t)$ changes so much faster than $f_N(t)$ that it can be a source of numerical instability; the calculation is very sensitive to $n$ and $t_0$. The choice of $R(t)$ in Eq.~\eqref{R(t)_2} gives good properties in the present case. Studies of process-dependent optimal forms of $R(t)$ could be interesting in the future.

\begin{figure*}
\begin{center}
\begin{tabular}{c}
\resizebox{175mm}{!}{\includegraphics{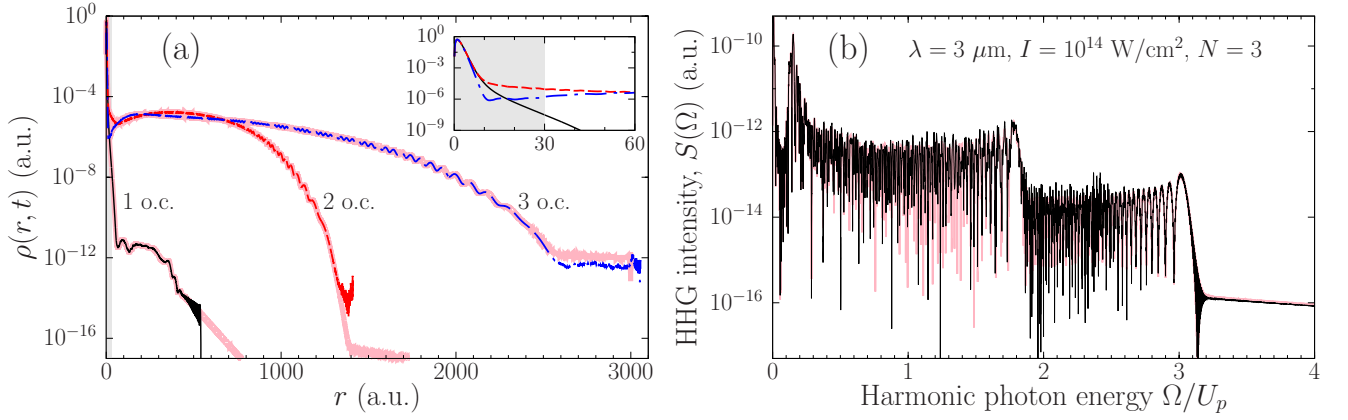}}
\end{tabular}
\caption{
\label{density_hhg_3um}
(Color online) Radial electron densities, $\rho(r,t)$ [Eq.~\eqref{Radial electron density}], at first, second, and third optical cycles (o.c.), i.e., at $t=(2\pi/\omega)\times n$ ($n=1,2$, and $3$). The inset shows the radial density around the origin; shaded area indicates the radial interval of the inner region ($0\le \xi(=r)\le r_{\Sigma}=30$). (b) HHG intensity, $S(\Omega)$, computed as a normed square of the Fourier transformation of dipole acceleration [Eq.~\eqref{S-ac}]. The harmonic photon energy is shown in units of the ponderomotive energy $U_p=F_0^2/(4\omega^2)=3.088$. Both (a) and (b) are computed in the velocity gauge for atomic hydrogen interacting with a three-cycle mid-IR laser pulse ($\lambda=3$ $\mu$m, $I=10^{14}$ W/cm$^2$, and $N=3$). The thin lines are the results of an ETS calculation, and the thick (pink) lines represent the ones of a non-time scaled calculation. See the main text for details.
}
\end{center}
\end{figure*}

\begin{figure}
\begin{center}
\begin{tabular}{c}
\resizebox{80mm}{!}{\includegraphics{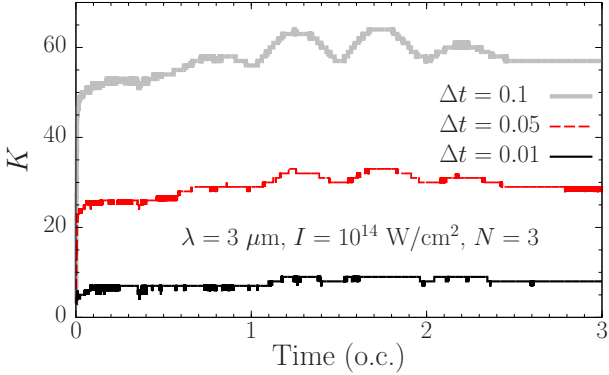}}
\end{tabular}
\caption{
\label{kry_dim_3um}
(Color online) Adapted dimension of the Krylov subspace, $K$, as a function of time [in units of optical cycle (o.c.)] in three ETS calculations using the same numerical parameters except the time step: $\Delta t=0.01$, $0.05$, and $0.1$. The value of $K$ is determined based on Eq.~\eqref{error 0} to ensure the error less than $\epsilon=10^{-15}$ at every time step. The calculations are carried out in the velocity gauge for atomic hydrogen interacting with a three-cycle mid-IR laser pulse ($\lambda=3$ $\mu$m, $I=10^{14}$ W/cm$^2$, and $N=3$). See the main text for details.
}
\end{center}
\end{figure}

\begin{figure*}
\begin{center}
\begin{tabular}{c}
\resizebox{175mm}{!}{\includegraphics{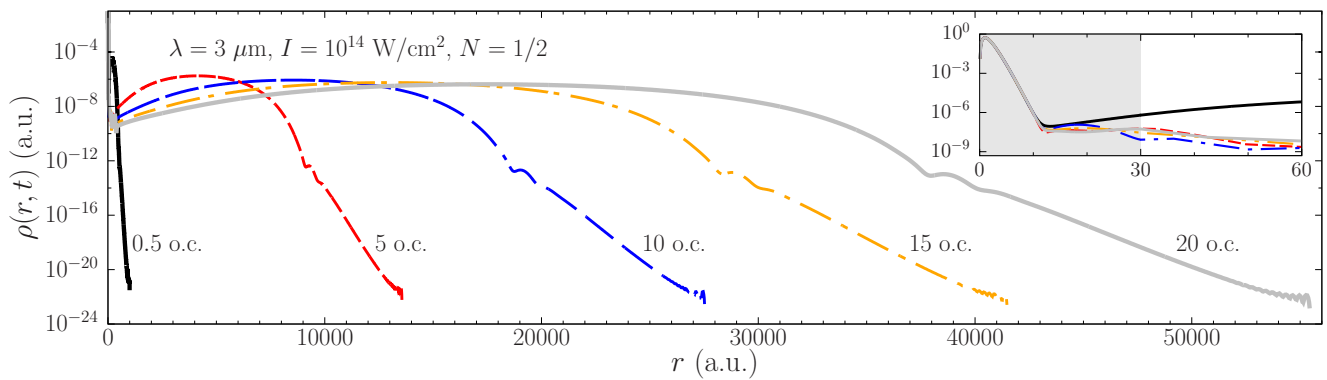}}
\end{tabular}
\caption{
\label{density_3um_h}
(Color online) Radial electron densities, $\rho(r,t)$ [Eq.~\eqref{Radial electron density}], at $0.5$th (just after the end of the pulse), $5$th, $10$th, $15$th, and $20$th optical cycles (o.c.), i.e., at $t=(2\pi/\omega)\times n$ ($n=0.5$, $5$, $10$, $15$, and $20$). The inset shows the radial density around the origin; shaded area indicates the radial interval of the inner region ($0\le \xi(=r)\le r_{\Sigma}=30$). These results are obtained by an ETS calculation in the length gauge for atomic hydrogen interacting with a half-cycle mid-IR laser pulse ($\lambda=3$, $\mu$m, $I=10^{14}$ W/cm$^2$, and $N=1/2$). After the completion of the pulse at $t=0.5$ o.c., the system time evolved for $19.5$ o.c. with no field. See the main text for details.
}
\end{center}
\end{figure*}

\begin{figure*}
\begin{center}
\begin{tabular}{c}
\resizebox{175mm}{!}{\includegraphics{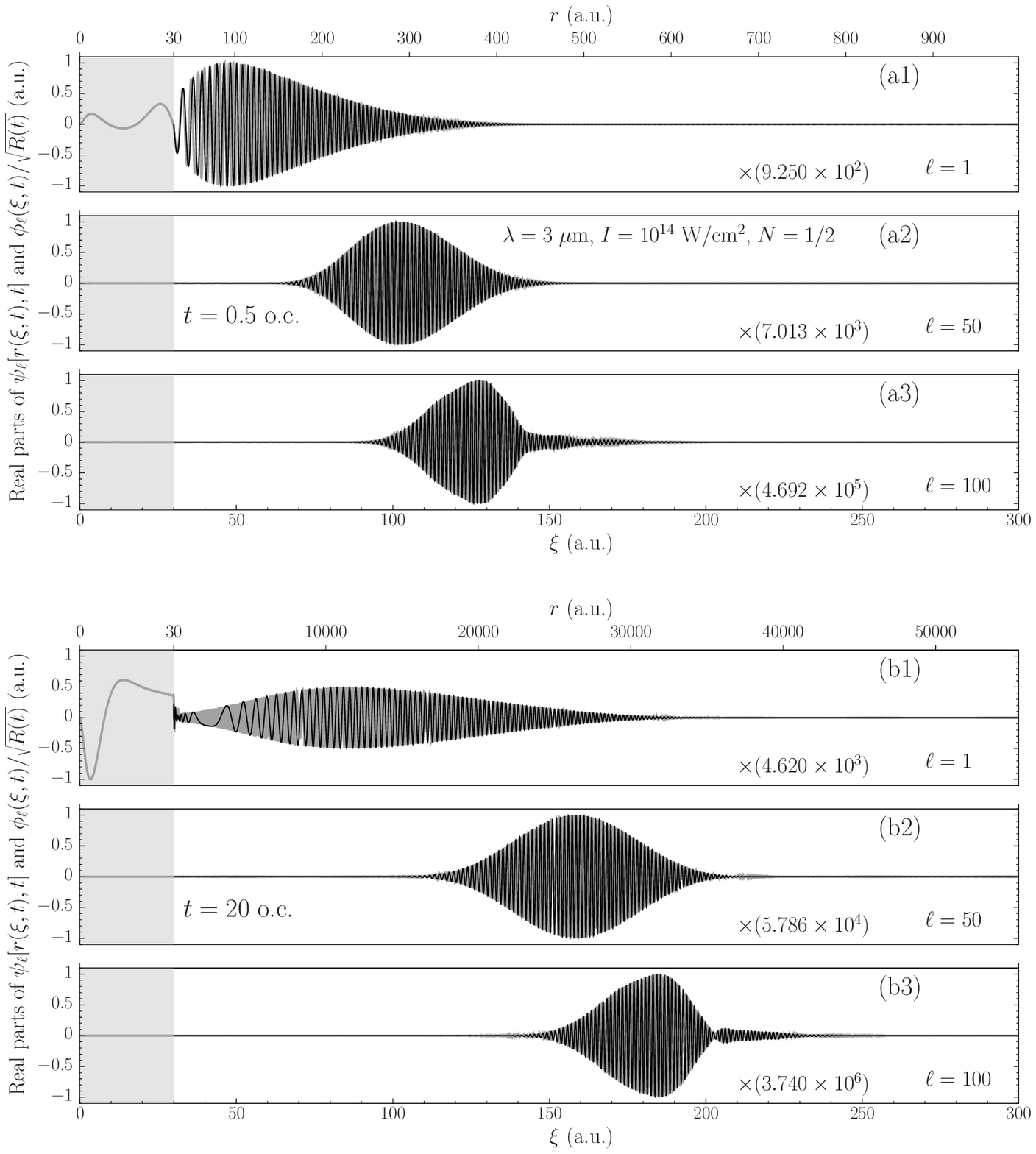}}
\end{tabular}
\caption{
\label{pwf_3um}
(a1--3) Real parts of $\psi_{\ell}[r(\xi,t),t]$ (thick dark-gray lines) and $\phi_{\ell}(\xi,t)/\sqrt{R(t)}$ (thin black lines) for $\ell=1$, $50$, and $100$ at $0.5$th optical cycles (o.c.); (b1--3) the same but at $20$th o.c. The shaded area in each panel indicates the radial interval of the inner region ($0\le \xi(=r)\le r_{\Sigma}=30$). For better visibility and comparison, each pair of curves is plotted after multiplication by the inverse of $\max \big|{\rm Re}[\psi_{\ell}[r(\xi,t),t]]\big|$ as indicated in each panel. The non-smooth appearance of ${\rm Re}[\psi_{\ell}[r(\xi,t),t]]$ around $r_{\Sigma}$ in (b1) is just due to the abrupt change of the scaling in the $r$ coordinate. The results in this figure and Fig.~\ref{density_3um_h} are obtained from the same ETS calculation in the length gauge for atomic hydrogen interacting with a half-cycle mid-IR laser pulse ($\lambda=3$ $\mu$m, $I=10^{14}$ W/cm$^2$, and $N=1/2$). See the main text for details.
}
\end{center}
\end{figure*}

\subsection{\label{HHG}Interaction with a three-cycle pulse}

Let us consider a three-cycle $(N=3)$ pulse the vector potential of which is given by
\begin{eqnarray}
A(t)=
\frac{F_0}{\omega}f_3(t)\sin\omega t,
\label{laser1}
\end{eqnarray}
where $F_0=\sqrt{I}=5.338\times10^{-2}$ with the intensity $I=10^{14}$ W/cm$^2=2.849\times10^{-3}$. Figure~\ref{fig_bound_states} depicts the profile of the electric field in the top-right corner. Figure~\ref{density_hhg_3um}(a) displays the radial electron densities of atomic hydrogen, 
\begin{eqnarray}
\rho(r,t)
=
\sum_{\ell=0}^{\ell_{\max}}
\big|\psi_{\ell}[\xi(r,t),t]\big|^2,
\label{Radial electron density}
\end{eqnarray}
computed in the velocity gauge by the ETS method and the usual non-time scaled method. The snapshot at each end of cycles ($t=(2\pi/\omega)\times n$, ($n=1,2,$ and 3)) exhibits the extension of the spatial radius in the ETS calculation, while the radius in the non-time scaled calculation keeps $3000$, constant. The inset of Fig.~\ref{density_hhg_3um}(a) displays the smooth continuity of the radial density across $r_{\Sigma}$, showing the stability of the ETS implementation. As a realistic observable, Fig.~\ref{density_hhg_3um}(b) displays the HHG spectrum computed as a normed square of the Fourier transformation of the dipole acceleration along the $z$ axis (polarization of the incoming pulse),
\begin{eqnarray}
\hspace{-4mm}  
S(\Omega)=
\Bigg|
\int^{T}_{0}
\langle\Psi(t)|
\left(
\frac{\partial}{\partial z}
\frac{1}{r}-F(t)
\right)
|\Psi(t)\rangle e^{i\Omega t}
dt
\Bigg|^2.
\label{S-ac}
\end{eqnarray}
The HHG spectrum consists of three plateaus: (1) $0<\Omega/U_p<0.1$, (2) $0.2<\Omega/U_p<1.8$, and (3) $1.9<\Omega/U_p<3.2$. Based upon the simple man's model (a simple classical simulation)~\cite{Corkum1993}, the electron trajectory characterized by the ionization and return times, $t_i^{(n)}$ and $t_r^{(n)}$, respectively, has the main responsibility to the $n$th plateau ($n=1,2$, and $3$): $2.03<t_i^{(1)}<2.20$ and $2.20<t_r^{(1)}<3.00$, $1.52<t_i^{(2)}<1.74$ and $1.74<t_r^{(2)}<2.77$, and $1.00<t_i^{(3)}<1.26$ and $1.26<t_r^{(3)}<1.98$ (in units of o.c.); also see the electric-field profile in the top-right corner of Fig.~\ref{fig_bound_states}. Such a clear time-to-energy mapping reflects the validity of the semiclassical picture of electron dynamics under the mid-IR lasers, and becomes less clear if $\lambda=3$ $\mu$m is replaced by $\lambda=0.8$ $\mu$m (not shown). A similar triple plateau in HHG spectra was recently reported for a different three-cycle pulse with $\lambda=1.6$ $\mu$m in Ref.~\cite{Shi2015}, where more detailed discussions are given based on the wavelet analysis with the help of the quantum orbit model.

In the ETS calculation for obtaining the results in Fig.~\ref{density_hhg_3um}, the wave function is parametrized as follows: $\ell_{\max}=200$, $r_{\Sigma}=30$, $\xi_{\max}=450$, $N_{\rm dvr}=10$, $\hat{N}_{\rm fe}=20$, $\check{N}_{\rm fe}=280$, $\Delta\xi=1.5$, and $R_{\infty}=0.01$. The non-time scaled calculation uses the same parameters except $\xi_{\max}=3000$, $\check{N}_{\rm fe}=1980$, and $R_{\infty}=0$. The parameters for the stiffness-free time propagation are common in both calculations: $\Delta t=0.05$, $\epsilon=10^{-15}$, $E_{\rm cut}=900$, and $\tilde{N}_{\rm fe}=10$ (i.e., $\tilde{N}=89$, which is the same as in Figs.~\ref{stiff}(b) and \ref{kry01}(b)). The dimension of the Krylov subspace is adapted at each time step based on Eq.~\eqref{error 0}. Figure~\ref{kry_dim_3um} shows $K$ as a function of time in three ETS calculations with different time steps: $\Delta t=0.01$, $0.05$, and $0.1$ (the rest of parameters are the same). In every case, $K$ increases with the start of the laser pulse, but keeps smaller than the estimated maximum in Fig.~\ref{kry01}(b). These three calculations exemplify the decrease of $(K_{\max}-1)/\Delta t$ as $\Delta t$ increases: $(9-1)/0.01=800$, $(33-1)/0.05=640$, and $(64-1)/0.1=630$, i.e., decreasing numerical cost as a function of $\Delta t$ [see Eq.~\eqref{error 5}]. That is, setting $\Delta t$ larger improves the numerical efficiency as long as it is much smaller than the characteristic time scale of external fields; in this case, $\Delta t \ll 2\pi/\omega=413.7$ should be satisfied. Also note that, without the stiffness removal, the time propagation is virtually infeasible because Eq.~\eqref{error 0} requires $K$ to be much greater than $1000$ (the blank area in Fig.~\ref{kry01}(a)). 

Due to the presence of $R(t)$, the numerical implementation of the ETS working equations~\eqref{TSCSP_in}, \eqref{TSCSP_out}, and \eqref{TSCSP_in_out_b} consumes an extra CPU time, which is absent in the usual non-time scaled calculation, for updating several matrix elements in the outer region [the block-diagonal part, shown by dashed (blue) lines in Fig.~\ref{fedvr_matrix}, at each of angular momentum blocks along $\ell=\ell'$]. However, this operation does not cause a major problem. In the above calculations, for instance, the time propagation by the non-time scaled method takes about six times more CPU time than by the ETS method. On the other hand, the number of nonzero elements of $\tilde{\bf H}(t)$ in the non-time scaled calculation is $123135659$, which is $5.6$ times larger than in the ETS calculation, $21987359$. The efficiency by ETS is almost proportional to the reduction of nonzero matrix elements. However, we also need to mention a weak point  of ETS; the ETS calculation is sensitive to the numerical condition, and finding a proper set of parameters, especially a proper value of $R_{\infty}$ [see Eq.~\eqref{R(t)_2}], is not straightforward. Assuming $r\sim kt$, i.e., a classical relation between the radial position and momentum of the photoelectron at large times, Eq.~\eqref{R(t)_2} then indicates $k\sim R_{\infty} \xi$~\cite{Sidky2000,Derbov2003,Roudnev2005,Serov2007,Serov2008,Hamido2011,Frapiccini2015}. Thus one may estimate $R_{\infty}=\sqrt{2\cdot 10U_p}/ \xi_{\max}=0.017$, where $U_p=F_0^2/(4\omega^2)=3.088$, and $10U_p$ is the maximum kinetic energy the photoelectron acquires by recollisions with the parent ion~\cite{Yang1993,Paulus1994}. The replacement of $R_{\infty}=0.01$ by $0.017$ in the above ETS calculation, however, gives some numerical instability; $R_{\infty}=0.017$ results in too fast growth of the $r$ coordinate, and hence requires a larger number of FEDVR functions for the accurate description of recollisions; this is numerically unfavorable. The numerical stability also depends on the value of $r_{\Sigma}$. Although Figs.~\ref{fig_bound_states}(b1--3) show smooth continuity of the Coulomb potential function and $\psi_{\rm 1s}[\xi(r),t]$ across $r_{\Sigma}=30$ and seemingly imply their stable numerical treatment, the accurate description of electron recollisions during the growth of the $r$ coordinate is not so simple because of the involvement of many excited states. 

Finally note that, just for computing HHG, ETS is not so useful; one can use a relatively small simulation volume in the non-time scaled calculation by employing a CAP or ECS, because only the electron dynamics around the nucleus is important to the dipole acceleration. The computation finishes with the completion of the pulse and does not require a long-time evolution any further. The true potential of ETS is the capability of long-time propagation without losing the norm of wave function as illustrated in the next subsection.

\subsection{\label{Tunneling}Interaction with a half-cycle pulse}
 
We now consider a half-cycle $(N=1/2)$ pulse defined by the electric field,
\begin{eqnarray}
F(t)=-F_0f_{1/2}(t)
\sin\omega t,
\label{laser2}
\end{eqnarray}
and track the long-time evolution of the tunnel-ejected electron after the completion of the pulse. Note that Eq.~\eqref{laser2} does not comply with the condition for realistic light fields, $\int_0^TF(t)dt=0$ (see, e.g., Refs.~\cite{Madsen2002,Gavrila2002}), and gives a non-vanishing vector potential at $t=0$ and $t>T$. Hence this artificial pulse necessitates employing the length gauge. Figure~\ref{density_3um_h} displays the time evolution of the radial electron density [Eq.~\eqref{Radial electron density}]. At $t=0.5$ o.c., i.e., just after the completion of the pulse, the electron wave packet occupies a volume of radius about $1000$. Due to its broad energy spectrum, the wave packet then spreads over a vast expanse of volume and reaches $55000$ at $t=20$ o.c. There is neither difficulty nor instability to continue the evolution as long time as one wishes. The inset of Fig.~\ref{density_3um_h} monitors the smooth radial density across $r_{\Sigma}=30$. The numerical condition in this ETS calculation is the same as the one shown in Sec.~\ref{HHG} except $\xi_{\max}=300$, $\check{N}_{\rm fe}=180$, $R_{\infty}=0.025$, and the employment of the length gauge. 

Based on the classical picture again, one may think of $R_{\infty}=\sqrt{2U_p}/ \xi_{\max}=0.0083$, because now there is no recollision. This estimation is, however, too little to take into account the broad energy spectrum of the wave packet, and it is safer to set $R_{\infty}$ a few times larger. In contrast to the three-cycle pulse in Sec.~\ref{HHG}, the calculation is not sensitive to $R_{\infty}$ and other parameters; the same converged result as in Fig.~\ref{density_3um_h} is also obtained more easily by setting $R_{\infty}$ larger than $0.025$ and employing smaller values for $\xi_{\max}$, $\hat{N}_{\rm fe}$, and $\check{N}_{\rm fe}$. Such an insensitivity to numerical condition is due to the absence of recollision in the half-cycle pulse. That is, for the analysis of long-time evolution of electron wave packets in circularly or near-circularly polarized mid-IR pulses, ETS will show its true potential without concerns of the numerical sensitivity to the parameters. 

Finally analysis of the radial function for each $\ell$ will be worthwhile for realizing how the ETS method enables keeping numerical stability for very long-time evolution. Figure~\ref{pwf_3um} shows the real parts of $\psi_{\ell}[r(\xi,t),t]$ and $\phi_{\ell}(\xi,t)/\sqrt{R(t)}$ (multiplied by a constant; see the caption) for $\ell=1$, $50$, and $100$. At $t=0.5$ o.c., although the spatial volume has not much extended yet, the oscillation of ${\rm Re}[\psi_{\ell}[r(\xi,t),t]]$ is already so fast that it is almost invisible. After a long-time evolution, the situation becomes worse because $\psi_{\ell}[r(\xi,t),t]$ spreads over a vast interval in the $r$ coordinate with increasing its phase gradient outward from the center of the wave packet~
\footnote{
To see a typical behavior of the phase gradient, for simplicity, let us consider the time evolution of a one-dimensional free electron prepared at $t=0$ in a normalized Gaussian wave packet having width $\Delta x$ and central momentum $k_0$:
\begin{eqnarray*}
&&\psi(x,t)
=
\sqrt{\frac{1}{\sqrt{2\pi}}\frac{\Delta x}{(\Delta x)^2+it/2}}
\\
&&
\hspace{12mm}
\times
\exp
\bigg[
ik_0x -ik_0^2t/2
-\frac{(x-k_0t)^2}{4[(\Delta x)^2+it/2]}
\bigg].
\end{eqnarray*}
The derivative of the phase with respect to $x$, $d\big[\arg \psi(x,t)\big]/dx=k_0+t(x-k_0t)/\big[4(\Delta x)^4+t^2\big]$, linearly increases with $x$, and is greater than $k_0$ for $x>k_0t$. Also see the discussion in Ref.~\cite{Sidky2000}.
}. 
Direct numerical treatment of $\psi_{\ell}(r,t)$ by the usual non-time scaled method is hence virtually infeasible. On the other hand, because the phase transformation in Eq.~\eqref{wave_function} cancels out the growing phase gradient (see the discussion in Ref.~\cite{Sidky2000}), ${\rm Re}[\phi_{\ell}(\xi,t)/\sqrt{R(t)}]$ exhibits not so fast oscillation in the $\xi$ coordinate even after a very long-time evolution. Figure~\ref{pwf_3um} shows the increasing difficulty of direct numerical treatment of $\psi_{\ell}(r,t)$ as it spreads without the help of the time scaling and the phase transformation. 

\section{\label{Conclusion}Conclusion and outlook}

Aiming at an efficient numerical treatment of tunneling ionization of atoms and molecules by mid-IR lasers, the ETS theory is formulated as a generalization of GTS. The working equations for numerical implementation are derived in terms of FEDVR basis functions. The key idea of ETS is to divide the spatial volume into two regions: a small spherical sphere around the nucleus and its outside, and then to carry out the time scaling only to the radial coordinates outside. As a result, the continuum part of the photoelectron wave packet is time scaled in the outer region and prevented from reflection. On the other hand, the bound-state part in the inner region is not time scaled and does not shrink toward the origin. Hence, ETS is less stiff than GTS. Furthermore, the stiffness-free FEDVR-based Lanczos algorithm is established to completely eliminate any stiffness for the treatment of long-wavelength lasers. The test calculations for atomic hydrogen interacting with linearly polarized mid-IR pulses demonstrate the capability of ETS and the stiffness-free time propagator. The method shows its true potential for the detailed analysis of wave-packet dynamics in non-recollision situations.

The ETS method and the stiffness-free time propagator can be flexibly used in several coordinate systems, e.g., in hyperspherical coordinates, as mentioned in Ref.~\cite{Frapiccini2015}, to treat photoionization of atomic helium. Application to RMT~\cite{Nikolopoulos2008,Moore2011a,Hart2014} may also be possible, but generalization to the TD-(RAS/GAS)CI method~\cite{Hochstuhl2012,Bauch2014} is most straightforward, enabling the extension to many-electron systems. Setting aside the many-electron problem, tunneling ionization of atoms and molecules by strong mid-IR lasers with arbitrary polarizations remains unexplored even within the SAE approximation. Toward this direction of research, in particular for circular or near-circular polarization, the potential of ETS is very promising. 

\begin{acknowledgments}
This work was supported by the ERC-StG (Project No. 277767-TDMET), and the VKR center of excellence, QUSCOPE. The numerical  results presented in this work were performed at the Centre for Scientific Computing, Aarhus http:/$\!$/phys.au.dk/forskning/cscaa/.
\end{acknowledgments}

\appendix*

\section{\label{Velocity gauge}Light-atom interaction operator in the velocity gauge}

The FEDVR-based working equations are derived in the length gauge in Sec.~\ref{ETS_FEDVR}. The following is a supplementary list of instructions needed to rewrite the light-atom interaction terms in Eqs.~\eqref{TSCSP_in}, \eqref{TSCSP_out}, and \eqref{TSCSP_in_out_b} in the velocity gauge.
\begin{widetext}
\begin{eqnarray}
{\rm In\; Eq.~\eqref{TSCSP_in}},
&&
W^{\rm L}_{\ell\ell'}(\xi_{\hat{\kappa}},t)
a_{\hat{\kappa}\ell'}(t)
\nonumber \\
\to&&
\int_0^{r_{\Sigma}}
\chi_{\hat{\kappa}}(\xi)
W^{\rm V}_{\ell\ell'}(\xi,\partial_{\xi},t\big) 
\psi_{\ell'}(\xi,t)
d\xi
\nonumber \\
&=&
-i\frac{A(t)}{2}g_{\ell\ell'}
\Bigg\{
\sum_{\hat{\kappa}'} 
a_{\hat{\kappa}'\ell'}(t)
\int_0^{\xi_{\max}}
\left[
\chi_{\hat{\kappa}}(\xi)
\frac{d\chi_{\hat{\kappa}'}(\xi)}{d\xi}
-
\frac{d\chi_{\hat{\kappa}}(\xi)}{d\xi}
\chi_{\hat{\kappa}'}(\xi)
\right]
d\xi
\nonumber \\
&&
+
\sqrt{\frac{2}{1+R(t)}}
a_{\ell'}(t)
\int_0^{\xi_{\max}}
\left[
\chi_{\hat{\kappa}}(\xi)
\frac{d\chi_{\rm b}(\xi)}{d\xi}
-
\frac{d\chi_{\hat{\kappa}}(\xi)}{d\xi}
\chi_{\rm b}(\xi)
\right]
d\xi
+
\frac{\ell'(\ell'+1)-\ell(\ell+1)}{\xi_{\hat{\kappa}}}
a_{\hat{\kappa}\ell'}(t)
\Bigg\}.
\label{TSCSP_in_velocity} 
\end{eqnarray}
\begin{eqnarray}
{\rm In\; Eq.~\eqref{TSCSP_out}},
&&
W^{\rm L}_{\ell\ell'}[r_{\Sigma}+R(t)(\xi_{\check{\kappa}}-r_{\Sigma}),t]
b_{\check{\kappa}\ell'}(t)
\nonumber \\
\to&&
\int_{r_{\Sigma}+0}^{\xi_{\max}}
\chi_{\check{\kappa}}(\xi)
\Big\{
W^{\rm V}_{\ell\ell'}\big[r_{\Sigma}+R(t)(\xi-r_{\Sigma}),\partial_{R(t)\xi},t\big]
+ g_{\ell\ell'} A(t)\dot{R}(t)(\xi-r_{\Sigma})
\Big\}
\phi_{\ell'}(\xi,t)
d\xi
\nonumber \\
&=&
-i\frac{A(t)}{2} g_{\ell\ell'}
\Bigg\{
\sum_{\check{\kappa}'} 
\frac{b_{\check{\kappa}'\ell'}(t)}{R(t)}
\int_0^{\xi_{\max}}
\left[
\chi_{\check{\kappa}}(\xi)
\frac{d\chi_{\check{\kappa}'}(\xi)}{d\xi}
-
\frac{d\chi_{\check{\kappa}}(\xi)}{d\xi}
\chi_{\check{\kappa}'}(\xi)
\right]
d\xi
\nonumber \\
&&
+
\sqrt{\frac{2R(t)}{1+R(t)}}
a_{\ell'}(t)
\int_0^{\xi_{\max}}
\left[
\chi_{\check{\kappa}}(\xi)
\frac{d\chi_{\rm b}(\xi)}{d\xi}
-
\frac{d\chi_{\check{\kappa}}(\xi)}{d\xi}
\chi_{\rm b}(\xi)
\right]
d\xi
+
\frac{\ell'(\ell'+1)-\ell(\ell+1)}{r_{\Sigma}+R(t)(\xi_{\check{\kappa}}-r_{\Sigma})}
b_{\check{\kappa}\ell'}(t)
\Bigg\}
\nonumber \\
&&
+
g_{\ell\ell'}
A(t)\dot{R}(t)(\xi_{\check{\kappa}}-r_{\Sigma})
b_{\check{\kappa}\ell}(t).
\label{TSCSP_out_velocity} 
\end{eqnarray}
\begin{eqnarray}
{\rm In\; Eq.~\eqref{TSCSP_in_out_b}},
&&
W^{\rm L}_{\ell\ell'}(r_{\Sigma},t)
a_{\ell'}(t)
\nonumber \\
\to&&
\sqrt{\frac{2}{1+R(t)}}
\Bigg[
\int_0^{r_{\Sigma}}
\chi_{\rm b}(\xi)
W^{\rm V}_{\ell\ell'}(\xi,\partial_{\xi},t\big) 
\psi_{\ell'}(\xi,t)
d\xi
\nonumber \\
&&
+
\sqrt{R(t)}
\int_{r_{\Sigma}+0}^{\xi_{\max}}
\chi_{\rm b}(\xi)
\Big\{
W^{\rm V}_{\ell\ell'}\big[r_{\Sigma}+R(t)(\xi-r_{\Sigma}),\partial_{R(t)\xi},t\big]
+ g_{\ell\ell'} A(t)\dot{R}(t)(\xi-r_{\Sigma})
\Big\}
\phi_{\ell'}(\xi,t)
d\xi
\Bigg]
\nonumber \\
&=&
-i\frac{A(t)}{2} g_{\ell\ell'}
\Bigg\{
\sqrt{\frac{2}{1+R(t)}}
\sum_{\hat{\kappa}} 
a_{\hat{\kappa}\ell'}(t)
\int_0^{\xi_{\max}}
\left[
\chi_{\rm b}(\xi)
\frac{d\chi_{\hat{\kappa}}(\xi)}{d\xi}
-
\frac{d\chi_{\rm b}(\xi)}{d\xi}
\chi_{\hat{\kappa}}(\xi)
\right]
d\xi
\nonumber \\
&&
+
\sqrt{\frac{2}{R(t)[1+R(t)]}}
\sum_{\check{\kappa}} 
b_{\check{\kappa}\ell'}(t)
\int_0^{\xi_{\max}}
\left[
\chi_{\rm b}(\xi)
\frac{d\chi_{\check{\kappa}}(\xi)}{d\xi}
-
\frac{d\chi_{\rm b}(\xi)}{d\xi}
\chi_{\check{\kappa}}(\xi)
\right]
d\xi
+
\frac{\ell'(\ell'+1)-\ell(\ell+1)}{r_{\Sigma}}
a_{\ell'}(t)
\Bigg\}.
\nonumber \\
\label{TSCSP_in_out_b_velocity} 
\end{eqnarray}
\end{widetext}

%


\begin{thebibliography}{61}%
\makeatletter
\providecommand \@ifxundefined [1]{%
 \@ifx{#1\undefined}
}%
\providecommand \@ifnum [1]{%
 \ifnum #1\expandafter \@firstoftwo
 \else \expandafter \@secondoftwo
 \fi
}%
\providecommand \@ifx [1]{%
 \ifx #1\expandafter \@firstoftwo
 \else \expandafter \@secondoftwo
 \fi
}%
\providecommand \natexlab [1]{#1}%
\providecommand \enquote  [1]{``#1''}%
\providecommand \bibnamefont  [1]{#1}%
\providecommand \bibfnamefont [1]{#1}%
\providecommand \citenamefont [1]{#1}%
\providecommand \href@noop [0]{\@secondoftwo}%
\providecommand \href [0]{\begingroup \@sanitize@url \@href}%
\providecommand \@href[1]{\@@startlink{#1}\@@href}%
\providecommand \@@href[1]{\endgroup#1\@@endlink}%
\providecommand \@sanitize@url [0]{\catcode `\\12\catcode `\$12\catcode
  `\&12\catcode `\#12\catcode `\^12\catcode `\_12\catcode `\%12\relax}%
\providecommand \@@startlink[1]{}%
\providecommand \@@endlink[0]{}%
\providecommand \url  [0]{\begingroup\@sanitize@url \@url }%
\providecommand \@url [1]{\endgroup\@href {#1}{\urlprefix }}%
\providecommand \urlprefix  [0]{URL }%
\providecommand \Eprint [0]{\href }%
\providecommand \doibase [0]{http://dx.doi.org/}%
\providecommand \selectlanguage [0]{\@gobble}%
\providecommand \bibinfo  [0]{\@secondoftwo}%
\providecommand \bibfield  [0]{\@secondoftwo}%
\providecommand \translation [1]{[#1]}%
\providecommand \BibitemOpen [0]{}%
\providecommand \bibitemStop [0]{}%
\providecommand \bibitemNoStop [0]{.\EOS\space}%
\providecommand \EOS [0]{\spacefactor3000\relax}%
\providecommand \BibitemShut  [1]{\csname bibitem#1\endcsname}%
\let\auto@bib@innerbib\@empty
\bibitem [{\citenamefont {Wolter}\ \emph {et~al.}(2015)\citenamefont {Wolter},
  \citenamefont {Pullen}, \citenamefont {Baudisch}, \citenamefont {Sclafani},
  \citenamefont {Hemmer}, \citenamefont {Senftleben}, \citenamefont
  {Schr{\"o}ter}, \citenamefont {Ullrich}, \citenamefont {Moshammerand},\ and\
  \citenamefont {Biegert}}]{Wolter2015}%
  \BibitemOpen
  \bibfield  {author} {\bibinfo {author} {\bibfnamefont {B.}~\bibnamefont
  {Wolter}}, \bibinfo {author} {\bibfnamefont {M.~G.}\ \bibnamefont {Pullen}},
  \bibinfo {author} {\bibfnamefont {M.}~\bibnamefont {Baudisch}}, \bibinfo
  {author} {\bibfnamefont {M.}~\bibnamefont {Sclafani}}, \bibinfo {author}
  {\bibfnamefont {M.}~\bibnamefont {Hemmer}}, \bibinfo {author} {\bibfnamefont
  {A.}~\bibnamefont {Senftleben}}, \bibinfo {author} {\bibfnamefont {C.~D.}\
  \bibnamefont {Schr{\"o}ter}}, \bibinfo {author} {\bibfnamefont
  {J.}~\bibnamefont {Ullrich}}, \bibinfo {author} {\bibfnamefont
  {R.}~\bibnamefont {Moshammerand}}, \ and\ \bibinfo {author} {\bibfnamefont
  {J.}~\bibnamefont {Biegert}},\ }\bibfield  {title} {\enquote {\bibinfo
  {title} {{Strong-Field Physics with Mid-IR Fields}},}\ }\href@noop {}
  {\bibfield  {journal} {\bibinfo  {journal} {Phys. Rev. X}\ }\textbf {\bibinfo
  {volume} {5}},\ \bibinfo {pages} {021034} (\bibinfo {year}
  {2015})}\BibitemShut {NoStop}%
\bibitem [{\citenamefont {Popmintchev}\ \emph {et~al.}(2012)\citenamefont
  {Popmintchev}, \citenamefont {Chen}, \citenamefont {Popmintchev},
  \citenamefont {Arpin}, \citenamefont {Brown}, \citenamefont
  {Ali{\v{s}}auskas}, \citenamefont {Andriukaitis}, \citenamefont
  {Bal{\v{c}}iunas}, \citenamefont {M{\"u}cke}, \citenamefont {Pugzlys},
  \citenamefont {Baltu{\v{s}}ska}, \citenamefont {Shim}, \citenamefont
  {Schrauth}, \citenamefont {Gaeta}, \citenamefont {Hern{\'a}ndez-Garc{\'i}a},
  \citenamefont {Plaja}, \citenamefont {Becker}, \citenamefont {Jaron-Becker},
  \citenamefont {Murnane},\ and\ \citenamefont {Kapteyn}}]{Popmintchev2012}%
  \BibitemOpen
  \bibfield  {author} {\bibinfo {author} {\bibfnamefont {T.}~\bibnamefont
  {Popmintchev}}, \bibinfo {author} {\bibfnamefont {M.-C.}\ \bibnamefont
  {Chen}}, \bibinfo {author} {\bibfnamefont {D.}~\bibnamefont {Popmintchev}},
  \bibinfo {author} {\bibfnamefont {P.}~\bibnamefont {Arpin}}, \bibinfo
  {author} {\bibfnamefont {S.}~\bibnamefont {Brown}}, \bibinfo {author}
  {\bibfnamefont {S.}~\bibnamefont {Ali{\v{s}}auskas}}, \bibinfo {author}
  {\bibfnamefont {G.}~\bibnamefont {Andriukaitis}}, \bibinfo {author}
  {\bibfnamefont {T.}~\bibnamefont {Bal{\v{c}}iunas}}, \bibinfo {author}
  {\bibfnamefont {O.~D.}\ \bibnamefont {M{\"u}cke}}, \bibinfo {author}
  {\bibfnamefont {A.}~\bibnamefont {Pugzlys}}, \bibinfo {author} {\bibfnamefont
  {A.}~\bibnamefont {Baltu{\v{s}}ska}}, \bibinfo {author} {\bibfnamefont
  {B.}~\bibnamefont {Shim}}, \bibinfo {author} {\bibfnamefont {S.~E.}\
  \bibnamefont {Schrauth}}, \bibinfo {author} {\bibfnamefont {A.}~\bibnamefont
  {Gaeta}}, \bibinfo {author} {\bibfnamefont {C.}~\bibnamefont
  {Hern{\'a}ndez-Garc{\'i}a}}, \bibinfo {author} {\bibfnamefont
  {L.}~\bibnamefont {Plaja}}, \bibinfo {author} {\bibfnamefont
  {A.}~\bibnamefont {Becker}}, \bibinfo {author} {\bibfnamefont
  {A.}~\bibnamefont {Jaron-Becker}}, \bibinfo {author} {\bibfnamefont {M.~M.}\
  \bibnamefont {Murnane}}, \ and\ \bibinfo {author} {\bibfnamefont {H.~C.}\
  \bibnamefont {Kapteyn}},\ }\bibfield  {title} {\enquote {\bibinfo {title}
  {{Bright Coherent Ultrahigh Harmonics in the keV X-ray Regime from
  Mid-Infrared Femtosecond Lasers}},}\ }\href@noop {} {\bibfield  {journal}
  {\bibinfo  {journal} {Science}\ }\textbf {\bibinfo {volume} {336}},\ \bibinfo
  {pages} {1287} (\bibinfo {year} {2012})}\BibitemShut {NoStop}%
\bibitem [{\citenamefont {Silva}\ \emph {et~al.}(2015)\citenamefont {Silva},
  \citenamefont {Teichmann}, \citenamefont {Cousin}, \citenamefont {Hemmer},\
  and\ \citenamefont {Biegert}}]{Silva2015}%
  \BibitemOpen
  \bibfield  {author} {\bibinfo {author} {\bibfnamefont {F.}~\bibnamefont
  {Silva}}, \bibinfo {author} {\bibfnamefont {S.~M.}\ \bibnamefont
  {Teichmann}}, \bibinfo {author} {\bibfnamefont {S.~L.}\ \bibnamefont
  {Cousin}}, \bibinfo {author} {\bibfnamefont {M.}~\bibnamefont {Hemmer}}, \
  and\ \bibinfo {author} {\bibfnamefont {J.}~\bibnamefont {Biegert}},\
  }\bibfield  {title} {\enquote {\bibinfo {title} {{Spatiotemporal isolation of
  attosecond soft X-ray pulses in the water window}},}\ }\href@noop {}
  {\bibfield  {journal} {\bibinfo  {journal} {Nat. Commun.}\ }\textbf {\bibinfo
  {volume} {6}},\ \bibinfo {pages} {6611} (\bibinfo {year} {2015})}\BibitemShut
  {NoStop}%
\bibitem [{\citenamefont {Hern{\'a}ndez-Garc{\'i}a}\ \emph
  {et~al.}(2013)\citenamefont {Hern{\'a}ndez-Garc{\'i}a}, \citenamefont
  {P{\'e}rez-Hern{\'a}ndez}, \citenamefont {Popmintchev}, \citenamefont
  {Murnane}, \citenamefont {Kapteyn}, \citenamefont {Jaron-Becker},
  \citenamefont {Becker},\ and\ \citenamefont {Plaja}}]{Garcia2013}%
  \BibitemOpen
  \bibfield  {author} {\bibinfo {author} {\bibfnamefont {C.}~\bibnamefont
  {Hern{\'a}ndez-Garc{\'i}a}}, \bibinfo {author} {\bibfnamefont {J.~A.}\
  \bibnamefont {P{\'e}rez-Hern{\'a}ndez}}, \bibinfo {author} {\bibfnamefont
  {T.}~\bibnamefont {Popmintchev}}, \bibinfo {author} {\bibfnamefont {M.~M.}\
  \bibnamefont {Murnane}}, \bibinfo {author} {\bibfnamefont {H.~C.}\
  \bibnamefont {Kapteyn}}, \bibinfo {author} {\bibfnamefont {A.}~\bibnamefont
  {Jaron-Becker}}, \bibinfo {author} {\bibfnamefont {A.}~\bibnamefont
  {Becker}}, \ and\ \bibinfo {author} {\bibfnamefont {L.}~\bibnamefont
  {Plaja}},\ }\bibfield  {title} {\enquote {\bibinfo {title} {{Zeptosecond High
  Harmonic keV X-Ray Waveforms Driven by Midinfrared Laser Pulses}},}\
  }\href@noop {} {\bibfield  {journal} {\bibinfo  {journal} {Phys. Rev. Lett.}\
  }\textbf {\bibinfo {volume} {111}},\ \bibinfo {pages} {033002} (\bibinfo
  {year} {2013})}\BibitemShut {NoStop}%
\bibitem [{\citenamefont {Tate}\ \emph {et~al.}(2007)\citenamefont {Tate},
  \citenamefont {Auguste}, \citenamefont {Muller}, \citenamefont
  {Sali{\'e}res}, \citenamefont {Agostini},\ and\ \citenamefont
  {DiMauro}}]{Tate2007}%
  \BibitemOpen
  \bibfield  {author} {\bibinfo {author} {\bibfnamefont {J.}~\bibnamefont
  {Tate}}, \bibinfo {author} {\bibfnamefont {T.}~\bibnamefont {Auguste}},
  \bibinfo {author} {\bibfnamefont {H.~G.}\ \bibnamefont {Muller}}, \bibinfo
  {author} {\bibfnamefont {P.}~\bibnamefont {Sali{\'e}res}}, \bibinfo {author}
  {\bibfnamefont {P.}~\bibnamefont {Agostini}}, \ and\ \bibinfo {author}
  {\bibfnamefont {L.~F.}\ \bibnamefont {DiMauro}},\ }\bibfield  {title}
  {\enquote {\bibinfo {title} {{Scaling of Wave-Packet Dynamics in an Intense
  Midinfrared Field}},}\ }\href@noop {} {\bibfield  {journal} {\bibinfo
  {journal} {Phys. Rev. Lett.}\ }\textbf {\bibinfo {volume} {98}},\ \bibinfo
  {pages} {013901} (\bibinfo {year} {2007})}\BibitemShut {NoStop}%
\bibitem [{\citenamefont {Huismans}\ \emph {et~al.}(2010)\citenamefont
  {Huismans}, \citenamefont {Rouz{\'e}e}, \citenamefont {Gijsbertsen},
  \citenamefont {Jungmann}, \citenamefont {Smolkowska}, \citenamefont {Logman},
  \citenamefont {L{\'e}pine}, \citenamefont {Cauchy}, \citenamefont {Zamith},
  \citenamefont {Marchenko}, \citenamefont {Bakker}, \citenamefont {Berden},
  \citenamefont {Redlich}, \citenamefont {van~der Meer}, \citenamefont
  {Muller}, \citenamefont {Vermin}, \citenamefont {Schafer}, \citenamefont
  {Spanner}, \citenamefont {Ivanov}, \citenamefont {Smirnova}, \citenamefont
  {Bauer}, \citenamefont {Popruzhenko},\ and\ \citenamefont
  {Vrakking}}]{Huismans2010}%
  \BibitemOpen
  \bibfield  {author} {\bibinfo {author} {\bibfnamefont {Y.}~\bibnamefont
  {Huismans}}, \bibinfo {author} {\bibfnamefont {A.}~\bibnamefont
  {Rouz{\'e}e}}, \bibinfo {author} {\bibfnamefont {A.}~\bibnamefont
  {Gijsbertsen}}, \bibinfo {author} {\bibfnamefont {J.~H.}\ \bibnamefont
  {Jungmann}}, \bibinfo {author} {\bibfnamefont {A.~S.}\ \bibnamefont
  {Smolkowska}}, \bibinfo {author} {\bibfnamefont {P.~S. W.~M.}\ \bibnamefont
  {Logman}}, \bibinfo {author} {\bibfnamefont {F.}~\bibnamefont {L{\'e}pine}},
  \bibinfo {author} {\bibfnamefont {C.}~\bibnamefont {Cauchy}}, \bibinfo
  {author} {\bibfnamefont {S.}~\bibnamefont {Zamith}}, \bibinfo {author}
  {\bibfnamefont {T.}~\bibnamefont {Marchenko}}, \bibinfo {author}
  {\bibfnamefont {J.~M.}\ \bibnamefont {Bakker}}, \bibinfo {author}
  {\bibfnamefont {G.}~\bibnamefont {Berden}}, \bibinfo {author} {\bibfnamefont
  {B.}~\bibnamefont {Redlich}}, \bibinfo {author} {\bibfnamefont {A.~F.~G.}\
  \bibnamefont {van~der Meer}}, \bibinfo {author} {\bibfnamefont {H.~G.}\
  \bibnamefont {Muller}}, \bibinfo {author} {\bibfnamefont {W.}~\bibnamefont
  {Vermin}}, \bibinfo {author} {\bibfnamefont {K.~J.}\ \bibnamefont {Schafer}},
  \bibinfo {author} {\bibfnamefont {M.}~\bibnamefont {Spanner}}, \bibinfo
  {author} {\bibfnamefont {M.~Yu.}\ \bibnamefont {Ivanov}}, \bibinfo {author}
  {\bibfnamefont {O.}~\bibnamefont {Smirnova}}, \bibinfo {author}
  {\bibfnamefont {D.}~\bibnamefont {Bauer}}, \bibinfo {author} {\bibfnamefont
  {S.~V.}\ \bibnamefont {Popruzhenko}}, \ and\ \bibinfo {author} {\bibfnamefont
  {M.~J.~J.}\ \bibnamefont {Vrakking}},\ }\bibfield  {title} {\enquote
  {\bibinfo {title} {{Time-Resolved Holography with Photoelectrons}},}\
  }\href@noop {} {\bibfield  {journal} {\bibinfo  {journal} {Science}\ }\textbf
  {\bibinfo {volume} {331}},\ \bibinfo {pages} {61} (\bibinfo {year}
  {2010})}\BibitemShut {NoStop}%
\bibitem [{\citenamefont {Huismans}\ \emph {et~al.}(2012)\citenamefont
  {Huismans}, \citenamefont {Gijsbertsen}, \citenamefont {Smolkowska},
  \citenamefont {Jungmann}, \citenamefont {Rouz{\'e}e}, \citenamefont {Logman},
  \citenamefont {Lz{\'e}pine}, \citenamefont {Cauchy}, \citenamefont {Zamith},
  \citenamefont {Marchenko}, \citenamefont {Bakker}, \citenamefont {Berden},
  \citenamefont {Redlich}, \citenamefont {van~der Meer}, \citenamefont
  {Ivanov}, \citenamefont {Yan}, \citenamefont {Bauer}, \citenamefont
  {Smirnova},\ and\ \citenamefont {Vrakking}}]{Huismans2012}%
  \BibitemOpen
  \bibfield  {author} {\bibinfo {author} {\bibfnamefont {Y.}~\bibnamefont
  {Huismans}}, \bibinfo {author} {\bibfnamefont {A.}~\bibnamefont
  {Gijsbertsen}}, \bibinfo {author} {\bibfnamefont {A.~S.}\ \bibnamefont
  {Smolkowska}}, \bibinfo {author} {\bibfnamefont {J.~H.}\ \bibnamefont
  {Jungmann}}, \bibinfo {author} {\bibfnamefont {A.}~\bibnamefont
  {Rouz{\'e}e}}, \bibinfo {author} {\bibfnamefont {P.~S. W.~M.}\ \bibnamefont
  {Logman}}, \bibinfo {author} {\bibfnamefont {F.}~\bibnamefont {Lz{\'e}pine}},
  \bibinfo {author} {\bibfnamefont {C.}~\bibnamefont {Cauchy}}, \bibinfo
  {author} {\bibfnamefont {S.}~\bibnamefont {Zamith}}, \bibinfo {author}
  {\bibfnamefont {T.}~\bibnamefont {Marchenko}}, \bibinfo {author}
  {\bibfnamefont {J.~M.}\ \bibnamefont {Bakker}}, \bibinfo {author}
  {\bibfnamefont {G.}~\bibnamefont {Berden}}, \bibinfo {author} {\bibfnamefont
  {B.}~\bibnamefont {Redlich}}, \bibinfo {author} {\bibfnamefont {A.~F.~G.}\
  \bibnamefont {van~der Meer}}, \bibinfo {author} {\bibfnamefont {M.~Yu.}\
  \bibnamefont {Ivanov}}, \bibinfo {author} {\bibfnamefont {T.-M.}\
  \bibnamefont {Yan}}, \bibinfo {author} {\bibfnamefont {D.}~\bibnamefont
  {Bauer}}, \bibinfo {author} {\bibfnamefont {O.}~\bibnamefont {Smirnova}}, \
  and\ \bibinfo {author} {\bibfnamefont {M.~J.~J.}\ \bibnamefont {Vrakking}},\
  }\bibfield  {title} {\enquote {\bibinfo {title} {{Scaling Laws for
  Photoelectron Holography in the Midinfrared Wavelength Regime}},}\
  }\href@noop {} {\bibfield  {journal} {\bibinfo  {journal} {Phys. Rev. Lett.}\
  }\textbf {\bibinfo {volume} {109}},\ \bibinfo {pages} {013002} (\bibinfo
  {year} {2012})}\BibitemShut {NoStop}%
\bibitem [{\citenamefont {Frolov}\ \emph {et~al.}(2015)\citenamefont {Frolov},
  \citenamefont {Manakov}, \citenamefont {Xiong}, \citenamefont {Peng},
  \citenamefont {Burgd{\"o}rfer},\ and\ \citenamefont {Starace}}]{Frolov2015}%
  \BibitemOpen
  \bibfield  {author} {\bibinfo {author} {\bibfnamefont {M.~V.}\ \bibnamefont
  {Frolov}}, \bibinfo {author} {\bibfnamefont {N.~L.}\ \bibnamefont {Manakov}},
  \bibinfo {author} {\bibfnamefont {W.-H.}\ \bibnamefont {Xiong}}, \bibinfo
  {author} {\bibfnamefont {L.-Y.}\ \bibnamefont {Peng}}, \bibinfo {author}
  {\bibfnamefont {J.}~\bibnamefont {Burgd{\"o}rfer}}, \ and\ \bibinfo {author}
  {\bibfnamefont {A.~F.}\ \bibnamefont {Starace}},\ }\bibfield  {title}
  {\enquote {\bibinfo {title} {{Scaling laws for high-order-harmonic generation
  with midinfrared laser pulses}},}\ }\href@noop {} {\bibfield  {journal}
  {\bibinfo  {journal} {Phys. Rev. A}\ }\textbf {\bibinfo {volume} {92}},\
  \bibinfo {pages} {023409} (\bibinfo {year} {2015})}\BibitemShut {NoStop}%
\bibitem [{\citenamefont {Tolstikhin}\ and\ \citenamefont
  {Morishita}(2012)}]{Tolstikhin2012}%
  \BibitemOpen
  \bibfield  {author} {\bibinfo {author} {\bibfnamefont {O.~I.}\ \bibnamefont
  {Tolstikhin}}\ and\ \bibinfo {author} {\bibfnamefont {T.}~\bibnamefont
  {Morishita}},\ }\bibfield  {title} {\enquote {\bibinfo {title} {{Adiabatic
  theory of ionization by intense laser pulses: Finite-range potentials}},}\
  }\href@noop {} {\bibfield  {journal} {\bibinfo  {journal} {Phys. Rev. A}\
  }\textbf {\bibinfo {volume} {86}},\ \bibinfo {pages} {043417} (\bibinfo
  {year} {2012})}\BibitemShut {NoStop}%
\bibitem [{\citenamefont {Ohmi}\ \emph {et~al.}(2015)\citenamefont {Ohmi},
  \citenamefont {Tolstikhin},\ and\ \citenamefont {Morishita}}]{Ohmi2015}%
  \BibitemOpen
  \bibfield  {author} {\bibinfo {author} {\bibfnamefont {M.}~\bibnamefont
  {Ohmi}}, \bibinfo {author} {\bibfnamefont {O.~I.}\ \bibnamefont
  {Tolstikhin}}, \ and\ \bibinfo {author} {\bibfnamefont {T.}~\bibnamefont
  {Morishita}},\ }\bibfield  {title} {\enquote {\bibinfo {title} {{Analysis of
  a shift of the maximum of photoelectron momentum distributions generated by
  intense circularly polarized pulses}},}\ }\href@noop {} {\bibfield  {journal}
  {\bibinfo  {journal} {Phys. Rev. A}\ }\textbf {\bibinfo {volume} {92}},\
  \bibinfo {pages} {043402} (\bibinfo {year} {2015})}\BibitemShut {NoStop}%
\bibitem [{\citenamefont {Eckle}\ \emph {et~al.}(2008)\citenamefont {Eckle},
  \citenamefont {Smolarski}, \citenamefont {Schlup}, \citenamefont {Biegert},
  \citenamefont {Staudte}, \citenamefont {Sch{\"o}ffler}, \citenamefont
  {Muller}, \citenamefont {D{\"o}rner},\ and\ \citenamefont
  {Keller}}]{Eckle2008}%
  \BibitemOpen
  \bibfield  {author} {\bibinfo {author} {\bibfnamefont {P.}~\bibnamefont
  {Eckle}}, \bibinfo {author} {\bibfnamefont {M.}~\bibnamefont {Smolarski}},
  \bibinfo {author} {\bibfnamefont {P.}~\bibnamefont {Schlup}}, \bibinfo
  {author} {\bibfnamefont {J.}~\bibnamefont {Biegert}}, \bibinfo {author}
  {\bibfnamefont {A.}~\bibnamefont {Staudte}}, \bibinfo {author} {\bibfnamefont
  {M.}~\bibnamefont {Sch{\"o}ffler}}, \bibinfo {author} {\bibfnamefont {H.~G.}\
  \bibnamefont {Muller}}, \bibinfo {author} {\bibfnamefont {R.}~\bibnamefont
  {D{\"o}rner}}, \ and\ \bibinfo {author} {\bibfnamefont {U.}~\bibnamefont
  {Keller}},\ }\bibfield  {title} {\enquote {\bibinfo {title} {{Attosecond
  angular streaking}},}\ }\href@noop {} {\bibfield  {journal} {\bibinfo
  {journal} {Nat. Phys}\ }\textbf {\bibinfo {volume} {4}},\ \bibinfo {pages}
  {565} (\bibinfo {year} {2008})}\BibitemShut {NoStop}%
\bibitem [{\citenamefont {Pfeiffer}\ \emph {et~al.}(2012)\citenamefont
  {Pfeiffer}, \citenamefont {Cirelli}, \citenamefont {Smolarski}, \citenamefont
  {Dimitrovski}, \citenamefont {Abu-samha}, \citenamefont {Madsen},\ and\
  \citenamefont {Keller}}]{Pfeiffer2012}%
  \BibitemOpen
  \bibfield  {author} {\bibinfo {author} {\bibfnamefont {A.~N.}\ \bibnamefont
  {Pfeiffer}}, \bibinfo {author} {\bibfnamefont {C.}~\bibnamefont {Cirelli}},
  \bibinfo {author} {\bibfnamefont {M.}~\bibnamefont {Smolarski}}, \bibinfo
  {author} {\bibfnamefont {D.}~\bibnamefont {Dimitrovski}}, \bibinfo {author}
  {\bibfnamefont {M.}~\bibnamefont {Abu-samha}}, \bibinfo {author}
  {\bibfnamefont {L.~B.}\ \bibnamefont {Madsen}}, \ and\ \bibinfo {author}
  {\bibfnamefont {U.}~\bibnamefont {Keller}},\ }\bibfield  {title} {\enquote
  {\bibinfo {title} {{Attoclock reveals natural coordinates of the
  laser-induced tunnelling current flow in atoms}},}\ }\href@noop {} {\bibfield
   {journal} {\bibinfo  {journal} {Nat. Phys.}\ }\textbf {\bibinfo {volume}
  {8}},\ \bibinfo {pages} {76} (\bibinfo {year} {2012})}\BibitemShut {NoStop}%
\bibitem [{\citenamefont {Petersen}\ \emph {et~al.}(2015)\citenamefont
  {Petersen}, \citenamefont {Henkel},\ and\ \citenamefont
  {Lein}}]{Petersen2015}%
  \BibitemOpen
  \bibfield  {author} {\bibinfo {author} {\bibfnamefont {I.}~\bibnamefont
  {Petersen}}, \bibinfo {author} {\bibfnamefont {J.}~\bibnamefont {Henkel}}, \
  and\ \bibinfo {author} {\bibfnamefont {M.}~\bibnamefont {Lein}},\ }\bibfield
  {title} {\enquote {\bibinfo {title} {{Signatures of Molecular Orbital
  Structure in Lateral Electron Momentum Distributions from Strong-Field
  Ionization}},}\ }\href@noop {} {\bibfield  {journal} {\bibinfo  {journal}
  {Phys. Rev. Lett.}\ }\textbf {\bibinfo {volume} {114}},\ \bibinfo {pages}
  {103004} (\bibinfo {year} {2015})}\BibitemShut {NoStop}%
\bibitem [{\citenamefont {Fleischer}\ \emph {et~al.}(2014)\citenamefont
  {Fleischer}, \citenamefont {Kfir}, \citenamefont {Diskin}, \citenamefont
  {Sidorenko},\ and\ \citenamefont {Cohen}}]{Fleischer2014}%
  \BibitemOpen
  \bibfield  {author} {\bibinfo {author} {\bibfnamefont {A.}~\bibnamefont
  {Fleischer}}, \bibinfo {author} {\bibfnamefont {O.}~\bibnamefont {Kfir}},
  \bibinfo {author} {\bibfnamefont {T.}~\bibnamefont {Diskin}}, \bibinfo
  {author} {\bibfnamefont {P.}~\bibnamefont {Sidorenko}}, \ and\ \bibinfo
  {author} {\bibfnamefont {O.}~\bibnamefont {Cohen}},\ }\bibfield  {title}
  {\enquote {\bibinfo {title} {{Spin angular momentum and tunable polarization
  in high-harmonic generation}},}\ }\href@noop {} {\bibfield  {journal}
  {\bibinfo  {journal} {Nat. Photonics}\ }\textbf {\bibinfo {volume} {8}},\
  \bibinfo {pages} {543} (\bibinfo {year} {2014})}\BibitemShut {NoStop}%
\bibitem [{\citenamefont {Mancuso}\ \emph {et~al.}(2015)\citenamefont
  {Mancuso}, \citenamefont {Hickstein}, \citenamefont {Grychtol}, \citenamefont
  {Knut}, \citenamefont {Kfir}, \citenamefont {Tong}, \citenamefont {Dollar},
  \citenamefont {Zusin}, \citenamefont {Gopalakrishnan}, \citenamefont
  {Gentry}, \citenamefont {Turgut}, \citenamefont {Ellis}, \citenamefont
  {Chen}, \citenamefont {Fleischer}, \citenamefont {Cohen}, \citenamefont
  {Kapteyn},\ and\ \citenamefont {Murnane}}]{Mancuso2015}%
  \BibitemOpen
  \bibfield  {author} {\bibinfo {author} {\bibfnamefont {C.~A.}\ \bibnamefont
  {Mancuso}}, \bibinfo {author} {\bibfnamefont {D.~D.}\ \bibnamefont
  {Hickstein}}, \bibinfo {author} {\bibfnamefont {P.}~\bibnamefont {Grychtol}},
  \bibinfo {author} {\bibfnamefont {R.}~\bibnamefont {Knut}}, \bibinfo {author}
  {\bibfnamefont {O.}~\bibnamefont {Kfir}}, \bibinfo {author} {\bibfnamefont
  {X.-M.}\ \bibnamefont {Tong}}, \bibinfo {author} {\bibfnamefont
  {F.}~\bibnamefont {Dollar}}, \bibinfo {author} {\bibfnamefont
  {D.}~\bibnamefont {Zusin}}, \bibinfo {author} {\bibfnamefont
  {M.}~\bibnamefont {Gopalakrishnan}}, \bibinfo {author} {\bibfnamefont
  {C.}~\bibnamefont {Gentry}}, \bibinfo {author} {\bibfnamefont
  {E.}~\bibnamefont {Turgut}}, \bibinfo {author} {\bibfnamefont {J.~L.}\
  \bibnamefont {Ellis}}, \bibinfo {author} {\bibfnamefont {Ming-Chang}\
  \bibnamefont {Chen}}, \bibinfo {author} {\bibfnamefont {A.}~\bibnamefont
  {Fleischer}}, \bibinfo {author} {\bibfnamefont {O.}~\bibnamefont {Cohen}},
  \bibinfo {author} {\bibfnamefont {H.~C.}\ \bibnamefont {Kapteyn}}, \ and\
  \bibinfo {author} {\bibfnamefont {M.~M.}\ \bibnamefont {Murnane}},\
  }\bibfield  {title} {\enquote {\bibinfo {title} {{Strong-field ionization
  with two-color circularly polarized laser fields}},}\ }\href@noop {}
  {\bibfield  {journal} {\bibinfo  {journal} {Phys. Rev. A}\ }\textbf {\bibinfo
  {volume} {91}},\ \bibinfo {pages} {031402(R)} (\bibinfo {year}
  {2015})}\BibitemShut {NoStop}%
\bibitem [{\citenamefont {Xie}(2015)}]{Xie2015}%
  \BibitemOpen
  \bibfield  {author} {\bibinfo {author} {\bibfnamefont {X.}~\bibnamefont
  {Xie}},\ }\bibfield  {title} {\enquote {\bibinfo {title} {{Two-Dimensional
  Attosecond Electron Wave-Packet Interferometry}},}\ }\href@noop {} {\bibfield
   {journal} {\bibinfo  {journal} {Phys. Rev. Lett.}\ }\textbf {\bibinfo
  {volume} {114}},\ \bibinfo {pages} {173003} (\bibinfo {year}
  {2015})}\BibitemShut {NoStop}%
\bibitem [{\citenamefont {Geng}\ \emph {et~al.}(2015)\citenamefont {Geng},
  \citenamefont {Xiong}, \citenamefont {Xiao}, \citenamefont {Peng},\ and\
  \citenamefont {Gong}}]{Geng2015}%
  \BibitemOpen
  \bibfield  {author} {\bibinfo {author} {\bibfnamefont {J.-W.}\ \bibnamefont
  {Geng}}, \bibinfo {author} {\bibfnamefont {W.-H.}\ \bibnamefont {Xiong}},
  \bibinfo {author} {\bibfnamefont {X.-R.}\ \bibnamefont {Xiao}}, \bibinfo
  {author} {\bibfnamefont {L.-Y.}\ \bibnamefont {Peng}}, \ and\ \bibinfo
  {author} {\bibfnamefont {Q.}~\bibnamefont {Gong}},\ }\bibfield  {title}
  {\enquote {\bibinfo {title} {{Nonadiabatic Electron Dynamics in Orthogonal
  Two-Color Laser Fields with Comparable Intensities}},}\ }\href@noop {}
  {\bibfield  {journal} {\bibinfo  {journal} {Phys. Rev. Lett.}\ }\textbf
  {\bibinfo {volume} {115}},\ \bibinfo {pages} {193001} (\bibinfo {year}
  {2015})}\BibitemShut {NoStop}%
\bibitem [{\citenamefont {Kulander}\ \emph {et~al.}(1992)\citenamefont
  {Kulander}, \citenamefont {Schafer},\ and\ \citenamefont
  {Krause}}]{Kulander1992}%
  \BibitemOpen
  \bibfield  {author} {\bibinfo {author} {\bibfnamefont {K.~C.}\ \bibnamefont
  {Kulander}}, \bibinfo {author} {\bibfnamefont {K.~J.}\ \bibnamefont
  {Schafer}}, \ and\ \bibinfo {author} {\bibfnamefont {J.~L.}\ \bibnamefont
  {Krause}},\ }\href@noop {} {\emph {\bibinfo {title} {{Atoms in Intense
  Radiation Fields}}}},\ edited by\ \bibinfo {editor} {\bibfnamefont
  {M.}~\bibnamefont {Gavrila}}\ (\bibinfo  {publisher} {Academic Press, New
  York},\ \bibinfo {year} {1992})\ pp.\ \bibinfo {pages} {247--300}\BibitemShut
  {NoStop}%
\bibitem [{\citenamefont {Pabst}\ and\ \citenamefont
  {Santra}(2013)}]{Pabst2013}%
  \BibitemOpen
  \bibfield  {author} {\bibinfo {author} {\bibfnamefont {S.}~\bibnamefont
  {Pabst}}\ and\ \bibinfo {author} {\bibfnamefont {R.}~\bibnamefont {Santra}},\
  }\bibfield  {title} {\enquote {\bibinfo {title} {{Strong-Field Many-Body
  Physics and the Giant Enhancement in the High-Harmonic Spectrum of Xenon}},}\
  }\href@noop {} {\bibfield  {journal} {\bibinfo  {journal} {Phys. Rev. Lett.}\
  }\textbf {\bibinfo {volume} {111}},\ \bibinfo {pages} {233005} (\bibinfo
  {year} {2013})}\BibitemShut {NoStop}%
\bibitem [{\citenamefont {Riss}\ and\ \citenamefont {Meyer}(1996)}]{Riss1996}%
  \BibitemOpen
  \bibfield  {author} {\bibinfo {author} {\bibfnamefont {U.~V.}\ \bibnamefont
  {Riss}}\ and\ \bibinfo {author} {\bibfnamefont {H.-D.}\ \bibnamefont
  {Meyer}},\ }\bibfield  {title} {\enquote {\bibinfo {title} {{Investigation on
  the reflection and transmission properties of complex absorbing
  potentials}},}\ }\href@noop {} {\bibfield  {journal} {\bibinfo  {journal} {J.
  Chem. Phys.}\ }\textbf {\bibinfo {volume} {105}},\ \bibinfo {pages} {1409}
  (\bibinfo {year} {1996})}\BibitemShut {NoStop}%
\bibitem [{\citenamefont {McCurdy}\ \emph {et~al.}(1991)\citenamefont
  {McCurdy}, \citenamefont {Stroud},\ and\ \citenamefont
  {Wisinski}}]{McCurdy1991}%
  \BibitemOpen
  \bibfield  {author} {\bibinfo {author} {\bibfnamefont {C.~W.}\ \bibnamefont
  {McCurdy}}, \bibinfo {author} {\bibfnamefont {C.~K.}\ \bibnamefont {Stroud}},
  \ and\ \bibinfo {author} {\bibfnamefont {M.~K.}\ \bibnamefont {Wisinski}},\
  }\bibfield  {title} {\enquote {\bibinfo {title} {{Solving the time-dependent
  {S}chr{\"o}dinger equation using complex-coordinate contours}},}\ }\href@noop
  {} {\bibfield  {journal} {\bibinfo  {journal} {Phys. Rev. A}\ }\textbf
  {\bibinfo {volume} {43}},\ \bibinfo {pages} {5980} (\bibinfo {year}
  {1991})}\BibitemShut {NoStop}%
\bibitem [{\citenamefont {He}\ \emph {et~al.}(2007)\citenamefont {He},
  \citenamefont {Ruiz},\ and\ \citenamefont {Becker}}]{He2007}%
  \BibitemOpen
  \bibfield  {author} {\bibinfo {author} {\bibfnamefont {F.}~\bibnamefont
  {He}}, \bibinfo {author} {\bibfnamefont {C.}~\bibnamefont {Ruiz}}, \ and\
  \bibinfo {author} {\bibfnamefont {A.}~\bibnamefont {Becker}},\ }\bibfield
  {title} {\enquote {\bibinfo {title} {{Absorbing boundaries in numerical
  solutions of the time-dependent Schr{\"o}dinger equation on a grid using
  exterior complex scaling}},}\ }\href@noop {} {\bibfield  {journal} {\bibinfo
  {journal} {Phys. Rev. A}\ }\textbf {\bibinfo {volume} {75}},\ \bibinfo
  {pages} {053407} (\bibinfo {year} {2007})}\BibitemShut {NoStop}%
\bibitem [{\citenamefont {Takemoto}\ \emph {et~al.}(2012)\citenamefont
  {Takemoto}, \citenamefont {Shimshovitz},\ and\ \citenamefont
  {Tannor}}]{Takemoto2012}%
  \BibitemOpen
  \bibfield  {author} {\bibinfo {author} {\bibfnamefont {N.}~\bibnamefont
  {Takemoto}}, \bibinfo {author} {\bibfnamefont {A.}~\bibnamefont
  {Shimshovitz}}, \ and\ \bibinfo {author} {\bibfnamefont {D.~J.}\ \bibnamefont
  {Tannor}},\ }\bibfield  {title} {\enquote {\bibinfo {title} {{Phase space
  approach to laser-driven electronic wavepacket propagation}},}\ }\href@noop
  {} {\bibfield  {journal} {\bibinfo  {journal} {J. Chem. Phys.}\ }\textbf
  {\bibinfo {volume} {137}},\ \bibinfo {pages} {011102} (\bibinfo {year}
  {2012})}\BibitemShut {NoStop}%
\bibitem [{\citenamefont {Soloviev}\ and\ \citenamefont
  {Vinitsky}(1985)}]{Solovievt1985}%
  \BibitemOpen
  \bibfield  {author} {\bibinfo {author} {\bibfnamefont {E.~A.}\ \bibnamefont
  {Soloviev}}\ and\ \bibinfo {author} {\bibfnamefont {S.~I.}\ \bibnamefont
  {Vinitsky}},\ }\bibfield  {title} {\enquote {\bibinfo {title} {{Suitable
  coordinates for the three-body problem in the adiabatic representation}},}\
  }\href@noop {} {\bibfield  {journal} {\bibinfo  {journal} {J. Phys. B}\
  }\textbf {\bibinfo {volume} {18}},\ \bibinfo {pages} {L557} (\bibinfo {year}
  {1985})}\BibitemShut {NoStop}%
\bibitem [{\citenamefont {Macek}\ \emph {et~al.}(1999)\citenamefont {Macek},
  \citenamefont {Ovchinnikov},\ and\ \citenamefont {Solov'ev}}]{Macek1999}%
  \BibitemOpen
  \bibfield  {author} {\bibinfo {author} {\bibfnamefont {J.~H.}\ \bibnamefont
  {Macek}}, \bibinfo {author} {\bibfnamefont {S.~Y.}\ \bibnamefont
  {Ovchinnikov}}, \ and\ \bibinfo {author} {\bibfnamefont {E.~A.}\ \bibnamefont
  {Solov'ev}},\ }\bibfield  {title} {\enquote {\bibinfo {title} {{Energy and
  angular distributions of detached electrons in a solvable model of ion-atom
  collisions}},}\ }\href@noop {} {\bibfield  {journal} {\bibinfo  {journal}
  {Phys. Rev. A}\ }\textbf {\bibinfo {volume} {60}},\ \bibinfo {pages} {1140}
  (\bibinfo {year} {1999})}\BibitemShut {NoStop}%
\bibitem [{\citenamefont {Sidky}\ and\ \citenamefont {Esry}(2000)}]{Sidky2000}%
  \BibitemOpen
  \bibfield  {author} {\bibinfo {author} {\bibfnamefont {E.~Y.}\ \bibnamefont
  {Sidky}}\ and\ \bibinfo {author} {\bibfnamefont {B.~D.}\ \bibnamefont
  {Esry}},\ }\bibfield  {title} {\enquote {\bibinfo {title} {{Boundary-Free
  Propagation with the Time-Dependent Schr{\"o}dinger Equation}},}\ }\href@noop
  {} {\bibfield  {journal} {\bibinfo  {journal} {Phys. Rev. Lett.}\ }\textbf
  {\bibinfo {volume} {85}},\ \bibinfo {pages} {5086} (\bibinfo {year}
  {2000})}\BibitemShut {NoStop}%
\bibitem [{\citenamefont {Derbov}\ \emph {et~al.}(2003)\citenamefont {Derbov},
  \citenamefont {Kaschiev}, \citenamefont {Serov}, \citenamefont {Gusev},\ and\
  \citenamefont {Vinitsky}}]{Derbov2003}%
  \BibitemOpen
  \bibfield  {author} {\bibinfo {author} {\bibfnamefont {V.~L.}\ \bibnamefont
  {Derbov}}, \bibinfo {author} {\bibfnamefont {M.~S.}\ \bibnamefont
  {Kaschiev}}, \bibinfo {author} {\bibfnamefont {V.~V.}\ \bibnamefont {Serov}},
  \bibinfo {author} {\bibfnamefont {A.~A.}\ \bibnamefont {Gusev}}, \ and\
  \bibinfo {author} {\bibfnamefont {S.~I.}\ \bibnamefont {Vinitsky}},\
  }\bibfield  {title} {\enquote {\bibinfo {title} {{Adaptive numerical methods
  for time-dependent Schr{\"o}dinger equation in atomic and laser physics}},}\
  }\href@noop {} {\bibfield  {journal} {\bibinfo  {journal} {Proc. SPIE}\
  }\textbf {\bibinfo {volume} {5067}},\ \bibinfo {pages} {218} (\bibinfo {year}
  {2003})}\BibitemShut {NoStop}%
\bibitem [{\citenamefont {Roudnev}\ and\ \citenamefont
  {Esry}(2005)}]{Roudnev2005}%
  \BibitemOpen
  \bibfield  {author} {\bibinfo {author} {\bibfnamefont {V.}~\bibnamefont
  {Roudnev}}\ and\ \bibinfo {author} {\bibfnamefont {B.~D.}\ \bibnamefont
  {Esry}},\ }\bibfield  {title} {\enquote {\bibinfo {title} {{HD$^+$
  photodissociation in the scaled coordinate approach}},}\ }\href@noop {}
  {\bibfield  {journal} {\bibinfo  {journal} {Phys. Rev. A}\ }\textbf {\bibinfo
  {volume} {71}},\ \bibinfo {pages} {013411} (\bibinfo {year}
  {2005})}\BibitemShut {NoStop}%
\bibitem [{\citenamefont {Serov}\ \emph {et~al.}(2007)\citenamefont {Serov},
  \citenamefont {Derbov}, \citenamefont {Joulakian},\ and\ \citenamefont
  {Vinitsky}}]{Serov2007}%
  \BibitemOpen
  \bibfield  {author} {\bibinfo {author} {\bibfnamefont {V.~V.}\ \bibnamefont
  {Serov}}, \bibinfo {author} {\bibfnamefont {V.~L.}\ \bibnamefont {Derbov}},
  \bibinfo {author} {\bibfnamefont {B.~B.}\ \bibnamefont {Joulakian}}, \ and\
  \bibinfo {author} {\bibfnamefont {S.~I.}\ \bibnamefont {Vinitsky}},\
  }\bibfield  {title} {\enquote {\bibinfo {title} {{Wave-packet-evolution
  approach for single and double ionization of two-electron systems by fast
  electrons}},}\ }\href@noop {} {\bibfield  {journal} {\bibinfo  {journal}
  {Phys. Rev. A}\ }\textbf {\bibinfo {volume} {75}},\ \bibinfo {pages} {012715}
  (\bibinfo {year} {2007})}\BibitemShut {NoStop}%
\bibitem [{\citenamefont {Serov}\ \emph {et~al.}(2008)\citenamefont {Serov},
  \citenamefont {Derbov}, \citenamefont {Joulakian},\ and\ \citenamefont
  {Vinitsky}}]{Serov2008}%
  \BibitemOpen
  \bibfield  {author} {\bibinfo {author} {\bibfnamefont {V.~V.}\ \bibnamefont
  {Serov}}, \bibinfo {author} {\bibfnamefont {V.~L.}\ \bibnamefont {Derbov}},
  \bibinfo {author} {\bibfnamefont {B.~B.}\ \bibnamefont {Joulakian}}, \ and\
  \bibinfo {author} {\bibfnamefont {S.~I.}\ \bibnamefont {Vinitsky}},\
  }\bibfield  {title} {\enquote {\bibinfo {title} {{Charge-scaling law for
  angular correlation in double photoionization of ions and atoms with two
  active electrons}},}\ }\href@noop {} {\bibfield  {journal} {\bibinfo
  {journal} {Phys. Rev. A}\ }\textbf {\bibinfo {volume} {78}},\ \bibinfo
  {pages} {063403} (\bibinfo {year} {2008})}\BibitemShut {NoStop}%
\bibitem [{\citenamefont {Hamido}\ \emph {et~al.}(2011)\citenamefont {Hamido},
  \citenamefont {Eiglsperger}, \citenamefont {Madro$\tilde{\rm n}$ero},
  \citenamefont {Mota-Furtado}, \citenamefont {O'Mahony}, \citenamefont
  {Frapiccini},\ and\ \citenamefont {Piraux}}]{Hamido2011}%
  \BibitemOpen
  \bibfield  {author} {\bibinfo {author} {\bibfnamefont {A.}~\bibnamefont
  {Hamido}}, \bibinfo {author} {\bibfnamefont {J.}~\bibnamefont {Eiglsperger}},
  \bibinfo {author} {\bibfnamefont {J.}~\bibnamefont {Madro$\tilde{\rm
  n}$ero}}, \bibinfo {author} {\bibfnamefont {F.}~\bibnamefont {Mota-Furtado}},
  \bibinfo {author} {\bibfnamefont {P.}~\bibnamefont {O'Mahony}}, \bibinfo
  {author} {\bibfnamefont {A.~L.}\ \bibnamefont {Frapiccini}}, \ and\ \bibinfo
  {author} {\bibfnamefont {B.}~\bibnamefont {Piraux}},\ }\bibfield  {title}
  {\enquote {\bibinfo {title} {{Time scaling with efficient time-propagation
  techniques for atoms and molecules in pulsed radiation fields}},}\
  }\href@noop {} {\bibfield  {journal} {\bibinfo  {journal} {Phys. Rev. A}\
  }\textbf {\bibinfo {volume} {84}},\ \bibinfo {pages} {013422} (\bibinfo
  {year} {2011})}\BibitemShut {NoStop}%
\bibitem [{\citenamefont {Frapiccini}\ \emph {et~al.}(2015)\citenamefont
  {Frapiccini}, \citenamefont {Hamido}, \citenamefont {Mota-Furtado},
  \citenamefont {O'Mahony},\ and\ \citenamefont {Piraux}}]{Frapiccini2015}%
  \BibitemOpen
  \bibfield  {author} {\bibinfo {author} {\bibfnamefont {A.~L.}\ \bibnamefont
  {Frapiccini}}, \bibinfo {author} {\bibfnamefont {A.}~\bibnamefont {Hamido}},
  \bibinfo {author} {\bibfnamefont {F.}~\bibnamefont {Mota-Furtado}}, \bibinfo
  {author} {\bibfnamefont {P.~F.}\ \bibnamefont {O'Mahony}}, \ and\ \bibinfo
  {author} {\bibfnamefont {B.}~\bibnamefont {Piraux}},\ }\bibfield  {title}
  {\enquote {\bibinfo {title} {{Multiresolution schemes for time-scaled
  propagation of wave packets}},}\ }\href@noop {} {\bibfield  {journal}
  {\bibinfo  {journal} {Phys. Rev. A}\ }\textbf {\bibinfo {volume} {91}},\
  \bibinfo {pages} {043423} (\bibinfo {year} {2015})}\BibitemShut {NoStop}%
\bibitem [{\citenamefont {Press}\ \emph {et~al.}(2007)\citenamefont {Press},
  \citenamefont {Teukolsky}, \citenamefont {Vetterling},\ and\ \citenamefont
  {Flannery}}]{Press2007}%
  \BibitemOpen
  \bibfield  {author} {\bibinfo {author} {\bibfnamefont {W.~H.}\ \bibnamefont
  {Press}}, \bibinfo {author} {\bibfnamefont {S.~A.}\ \bibnamefont
  {Teukolsky}}, \bibinfo {author} {\bibfnamefont {W.~T.}\ \bibnamefont
  {Vetterling}}, \ and\ \bibinfo {author} {\bibfnamefont {B.~P.}\ \bibnamefont
  {Flannery}},\ }\href@noop {} {\emph {\bibinfo {title} {Numerical Recipes: The
  Art of Scientific Computing}}},\ \bibinfo {edition} {3rd}\ ed.\ (\bibinfo
  {publisher} {Cambridge University Press},\ \bibinfo {address} {Cambridge},\
  \bibinfo {year} {2007})\BibitemShut {NoStop}%
\bibitem [{\citenamefont {Fatunla}(1978)}]{Fatunla1978}%
  \BibitemOpen
  \bibfield  {author} {\bibinfo {author} {\bibfnamefont {S.~O.}\ \bibnamefont
  {Fatunla}},\ }\bibfield  {title} {\enquote {\bibinfo {title} {{An Implicit
  Two-Point Numerical Integration Formula for Linear and Nonlinear Stiff
  Systems of Ordinary Differential Equations}},}\ }\href@noop {} {\bibfield
  {journal} {\bibinfo  {journal} {Math. Comput.}\ }\textbf {\bibinfo {volume}
  {32}},\ \bibinfo {pages} {1} (\bibinfo {year} {1978})}\BibitemShut {NoStop}%
\bibitem [{\citenamefont {Fatunla}(1980)}]{Fatunla1980}%
  \BibitemOpen
  \bibfield  {author} {\bibinfo {author} {\bibfnamefont {S.~O.}\ \bibnamefont
  {Fatunla}},\ }\bibfield  {title} {\enquote {\bibinfo {title} {{Numerical
  Integrators for Stiff and Highly Oscillatory Differential Equations}},}\
  }\href@noop {} {\bibfield  {journal} {\bibinfo  {journal} {Math. Comput.}\
  }\textbf {\bibinfo {volume} {34}},\ \bibinfo {pages} {373} (\bibinfo {year}
  {1980})}\BibitemShut {NoStop}%
\bibitem [{\citenamefont {Rescigno}\ and\ \citenamefont
  {McCurdy}(2000)}]{Rescigno2000}%
  \BibitemOpen
  \bibfield  {author} {\bibinfo {author} {\bibfnamefont {T.~N.}\ \bibnamefont
  {Rescigno}}\ and\ \bibinfo {author} {\bibfnamefont {C.~W.}\ \bibnamefont
  {McCurdy}},\ }\bibfield  {title} {\enquote {\bibinfo {title} {Numerical grid
  methods for quantum-mechanical scattering problems},}\ }\href@noop {}
  {\bibfield  {journal} {\bibinfo  {journal} {Phys. Rev. A}\ }\textbf {\bibinfo
  {volume} {62}},\ \bibinfo {pages} {032706} (\bibinfo {year}
  {2000})}\BibitemShut {NoStop}%
\bibitem [{\citenamefont {Park}\ and\ \citenamefont {Light}(1986)}]{Park1986}%
  \BibitemOpen
  \bibfield  {author} {\bibinfo {author} {\bibfnamefont {T.~J.}\ \bibnamefont
  {Park}}\ and\ \bibinfo {author} {\bibfnamefont {J.~C.}\ \bibnamefont
  {Light}},\ }\bibfield  {title} {\enquote {\bibinfo {title} {{Unitary quantum
  time evolution by iterative Lanczos reduction}},}\ }\href@noop {} {\bibfield
  {journal} {\bibinfo  {journal} {J. Chem. Phys.}\ }\textbf {\bibinfo {volume}
  {85}},\ \bibinfo {pages} {5870} (\bibinfo {year} {1986})}\BibitemShut
  {NoStop}%
\bibitem [{\citenamefont {Hochstuhl}\ and\ \citenamefont
  {Bonitz}(2012)}]{Hochstuhl2012}%
  \BibitemOpen
  \bibfield  {author} {\bibinfo {author} {\bibfnamefont {D.}~\bibnamefont
  {Hochstuhl}}\ and\ \bibinfo {author} {\bibfnamefont {M.}~\bibnamefont
  {Bonitz}},\ }\bibfield  {title} {\enquote {\bibinfo {title} {{Time-dependent
  restricted-active-space configuration-interaction method for the
  photoionization of many-electron atoms}},}\ }\href@noop {} {\bibfield
  {journal} {\bibinfo  {journal} {Phys. Rev. A}\ }\textbf {\bibinfo {volume}
  {86}},\ \bibinfo {pages} {053424} (\bibinfo {year} {2012})}\BibitemShut
  {NoStop}%
\bibitem [{\citenamefont {Bauch}\ \emph {et~al.}(2014)\citenamefont {Bauch},
  \citenamefont {S{\o}rensen},\ and\ \citenamefont {Madsen}}]{Bauch2014}%
  \BibitemOpen
  \bibfield  {author} {\bibinfo {author} {\bibfnamefont {S.}~\bibnamefont
  {Bauch}}, \bibinfo {author} {\bibfnamefont {L.~K.}\ \bibnamefont
  {S{\o}rensen}}, \ and\ \bibinfo {author} {\bibfnamefont {L.~B.}\ \bibnamefont
  {Madsen}},\ }\bibfield  {title} {\enquote {\bibinfo {title} {Time-dependent
  generalized-active-space configuration-interaction approach to
  photoionization dynamics of atoms and molecules},}\ }\href@noop {} {\bibfield
   {journal} {\bibinfo  {journal} {Phys. Rev. A}\ }\textbf {\bibinfo {volume}
  {90}},\ \bibinfo {pages} {062508} (\bibinfo {year} {2014})}\BibitemShut
  {NoStop}%
\bibitem [{\citenamefont {Nikolopoulos}\ \emph {et~al.}(2008)\citenamefont
  {Nikolopoulos}, \citenamefont {Parker},\ and\ \citenamefont
  {Taylor}}]{Nikolopoulos2008}%
  \BibitemOpen
  \bibfield  {author} {\bibinfo {author} {\bibfnamefont {L.~A.~A.}\
  \bibnamefont {Nikolopoulos}}, \bibinfo {author} {\bibfnamefont {J.~S.}\
  \bibnamefont {Parker}}, \ and\ \bibinfo {author} {\bibfnamefont {K.~T.}\
  \bibnamefont {Taylor}},\ }\bibfield  {title} {\enquote {\bibinfo {title}
  {{Combined {$R$}-matrix eigenstate basis set and finite-difference
  propagation method for the time-dependent {Schr{\"o}dinger} equation: {T}he
  one-electron case}},}\ }\href@noop {} {\bibfield  {journal} {\bibinfo
  {journal} {Phys. Rev. A}\ }\textbf {\bibinfo {volume} {78}},\ \bibinfo
  {pages} {063420} (\bibinfo {year} {2008})}\BibitemShut {NoStop}%
\bibitem [{\citenamefont {Moore}\ \emph {et~al.}(2011)\citenamefont {Moore},
  \citenamefont {Lysaght}, \citenamefont {Nikolopoulos}, \citenamefont
  {Parker}, \citenamefont {van~der Hart},\ and\ \citenamefont
  {Taylor}}]{Moore2011a}%
  \BibitemOpen
  \bibfield  {author} {\bibinfo {author} {\bibfnamefont {L.~R.}\ \bibnamefont
  {Moore}}, \bibinfo {author} {\bibfnamefont {M.~A.}\ \bibnamefont {Lysaght}},
  \bibinfo {author} {\bibfnamefont {L.~A.~A.}\ \bibnamefont {Nikolopoulos}},
  \bibinfo {author} {\bibfnamefont {J.~S.}\ \bibnamefont {Parker}}, \bibinfo
  {author} {\bibfnamefont {H.~W.}\ \bibnamefont {van~der Hart}}, \ and\
  \bibinfo {author} {\bibfnamefont {K.~T.}\ \bibnamefont {Taylor}},\ }\bibfield
   {title} {\enquote {\bibinfo {title} {The {RMT} method for many-electron
  atomic systems in intense short-pulse laser light},}\ }\href@noop {}
  {\bibfield  {journal} {\bibinfo  {journal} {J. Mod. Opt.}\ }\textbf {\bibinfo
  {volume} {58}},\ \bibinfo {pages} {1132} (\bibinfo {year}
  {2011})}\BibitemShut {NoStop}%
\bibitem [{\citenamefont {van~der Hart}(2014)}]{Hart2014}%
  \BibitemOpen
  \bibfield  {author} {\bibinfo {author} {\bibfnamefont {H.~W.}\ \bibnamefont
  {van~der Hart}},\ }\bibfield  {title} {\enquote {\bibinfo {title}
  {{Time-dependent $R$-matrix theory applied to two-photon double}},}\
  }\href@noop {} {\bibfield  {journal} {\bibinfo  {journal} {Phys. Rev. A}\
  }\textbf {\bibinfo {volume} {89}},\ \bibinfo {pages} {053407} (\bibinfo
  {year} {2014})}\BibitemShut {NoStop}%
\bibitem [{\citenamefont {Light}\ \emph {et~al.}(1985)\citenamefont {Light},
  \citenamefont {Hamilton},\ and\ \citenamefont {Lill}}]{Light1985}%
  \BibitemOpen
  \bibfield  {author} {\bibinfo {author} {\bibfnamefont {J.~C.}\ \bibnamefont
  {Light}}, \bibinfo {author} {\bibfnamefont {I.~P.}\ \bibnamefont {Hamilton}},
  \ and\ \bibinfo {author} {\bibfnamefont {J.~V.}\ \bibnamefont {Lill}},\
  }\bibfield  {title} {\enquote {\bibinfo {title} {Generalized discrete
  variable approximation in quantum mechanics},}\ }\href@noop {} {\bibfield
  {journal} {\bibinfo  {journal} {J. Chem. Phys.}\ }\textbf {\bibinfo {volume}
  {82}},\ \bibinfo {pages} {1400} (\bibinfo {year} {1985})}\BibitemShut
  {NoStop}%
\bibitem [{\citenamefont {Descouvemont}\ and\ \citenamefont
  {Baye}(2010)}]{Descouvemont2010}%
  \BibitemOpen
  \bibfield  {author} {\bibinfo {author} {\bibfnamefont {P.}~\bibnamefont
  {Descouvemont}}\ and\ \bibinfo {author} {\bibfnamefont {D.}~\bibnamefont
  {Baye}},\ }\bibfield  {title} {\enquote {\bibinfo {title} {The ${R}$-matrix
  theory},}\ }\href@noop {} {\bibfield  {journal} {\bibinfo  {journal} {Rep.
  Prog. Phys.}\ }\textbf {\bibinfo {volume} {73}},\ \bibinfo {pages} {036301}
  (\bibinfo {year} {2010})}\BibitemShut {NoStop}%
\bibitem [{\citenamefont {Burke}(2011)}]{Burke2011}%
  \BibitemOpen
  \bibfield  {author} {\bibinfo {author} {\bibfnamefont {P.~G.}\ \bibnamefont
  {Burke}},\ }\href@noop {} {\emph {\bibinfo {title} {${R}$-Matrix Theory of
  Atomic Collisions}}}\ (\bibinfo  {publisher} {Springer Verlag, Heidelberg},\
  \bibinfo {year} {2011})\BibitemShut {NoStop}%
\bibitem [{\citenamefont {Kuleff}\ \emph {et~al.}(2005)\citenamefont {Kuleff},
  \citenamefont {Breidbach},\ and\ \citenamefont {Cederbaum}}]{Kuleff2005}%
  \BibitemOpen
  \bibfield  {author} {\bibinfo {author} {\bibfnamefont {A.~I.}\ \bibnamefont
  {Kuleff}}, \bibinfo {author} {\bibfnamefont {J.}~\bibnamefont {Breidbach}}, \
  and\ \bibinfo {author} {\bibfnamefont {L.~S.}\ \bibnamefont {Cederbaum}},\
  }\bibfield  {title} {\enquote {\bibinfo {title} {{Multielectron wave-packet
  propagation: General theory and application}},}\ }\href@noop {} {\bibfield
  {journal} {\bibinfo  {journal} {J. Chem . Phys.}\ }\textbf {\bibinfo {volume}
  {123}},\ \bibinfo {pages} {044111} (\bibinfo {year} {2005})}\BibitemShut
  {NoStop}%
\bibitem [{\citenamefont {Arnoldi}(1951)}]{Arnoldi1951}%
  \BibitemOpen
  \bibfield  {author} {\bibinfo {author} {\bibfnamefont {W.~E.}\ \bibnamefont
  {Arnoldi}},\ }\bibfield  {title} {\enquote {\bibinfo {title} {{THE PRINCIPLE
  OF MINIMIZED ITERATIONS IN THE SOLUTION OF THE MATRIX EIGENVALUE PROBLEM}},}\
  }\href@noop {} {\bibfield  {journal} {\bibinfo  {journal} {Quart. Appl.
  Math.}\ }\textbf {\bibinfo {volume} {9}},\ \bibinfo {pages} {17} (\bibinfo
  {year} {1951})}\BibitemShut {NoStop}%
\bibitem [{Note1()}]{Note1}%
  \BibitemOpen
  \bibinfo {note} {For the time-independent Hamiltonian, i.e., if there is no
  light field so that ${\protect \bf H}(t)={\protect \bf H}(0)$, it can be
  shown that $B< \protect \sqrt [K-1]{2} \protect \qopname \relax
  m{max}\protect \{E_{n\ell }\protect \}_{n\ell } \protect \qopname \relax
  o{exp}{\setbox \z@ \hbox {\frozen@everymath \@emptytoks \mathsurround \z@
  $\nulldelimiterspace \z@ \left [\vcenter to\@ne \big@size {}\right .$}\box
  \z@ }{\protect \qopname \relax m{max}\protect \{E_{n\ell }\protect \}_{n\ell
  }/(K-1)}{\setbox \z@ \hbox {\frozen@everymath \@emptytoks \mathsurround \z@
  $\nulldelimiterspace \z@ \left ]\vcenter to\@ne \big@size {}\right .$}\box
  \z@ }$. Additionally, if ${\protect \bf H}(0)$ is positive definite, i.e., if
  ${\protect \bf H}(0)$ holds no bound state, it can also be shown that $B<
  2^{(K-2)/(K-1)} \protect \qopname \relax m{max}\protect \{E_{n\ell }\protect
  \}_{n\ell }$. The latter is not the case in our discussion, but may give
  better estimation to our error and stiffness assessment, because the
  stiffness mainly arises from very large angular momentum states supporting
  almost no bound state. See Refs.~\cite {Gallopoulos1992,Saad1992} for
  mathematical details.}\BibitemShut {Stop}%
\bibitem [{\citenamefont {Abramowitz}\ and\ \citenamefont
  {Stegun}(1972)}]{Abramowitz1972}%
  \BibitemOpen
  \bibinfo {editor} {\bibfnamefont {M.}~\bibnamefont {Abramowitz}}\ and\
  \bibinfo {editor} {\bibfnamefont {I.~A.}\ \bibnamefont {Stegun}},\ eds.,\
  \href@noop {} {\emph {\bibinfo {title} {Handbook of Mathematical
  Functions}}}\ (\bibinfo  {publisher} {Dover, New York},\ \bibinfo {year}
  {1972})\BibitemShut {NoStop}%
\bibitem [{\citenamefont {Guan}\ \emph {et~al.}(2008)\citenamefont {Guan},
  \citenamefont {Noble}, \citenamefont {Zatsarinny}, \citenamefont
  {Bartschat},\ and\ \citenamefont {Schneider}}]{Guan2008}%
  \BibitemOpen
  \bibfield  {author} {\bibinfo {author} {\bibfnamefont {X.}~\bibnamefont
  {Guan}}, \bibinfo {author} {\bibfnamefont {C.~J.}\ \bibnamefont {Noble}},
  \bibinfo {author} {\bibfnamefont {O.}~\bibnamefont {Zatsarinny}}, \bibinfo
  {author} {\bibfnamefont {K.}~\bibnamefont {Bartschat}}, \ and\ \bibinfo
  {author} {\bibfnamefont {B.~I.}\ \bibnamefont {Schneider}},\ }\bibfield
  {title} {\enquote {\bibinfo {title} {Time-dependent {$R$}-matrix calculations
  for multiphoton ionization of argon atoms in strong laser pulses},}\
  }\href@noop {} {\bibfield  {journal} {\bibinfo  {journal} {Phys. Rev. A}\
  }\textbf {\bibinfo {volume} {78}},\ \bibinfo {pages} {053402} (\bibinfo
  {year} {2008})}\BibitemShut {NoStop}%
\bibitem [{\citenamefont {Hochstuhl}\ \emph {et~al.}(2014)\citenamefont
  {Hochstuhl}, \citenamefont {Hinz},\ and\ \citenamefont
  {Bonitz}}]{Hochstuhl2014}%
  \BibitemOpen
  \bibfield  {author} {\bibinfo {author} {\bibfnamefont {D.}~\bibnamefont
  {Hochstuhl}}, \bibinfo {author} {\bibfnamefont {C.~M.}\ \bibnamefont {Hinz}},
  \ and\ \bibinfo {author} {\bibfnamefont {M.}~\bibnamefont {Bonitz}},\
  }\bibfield  {title} {\enquote {\bibinfo {title} {{Time-dependent
  multiconfiguration methods for the numerical simulation of photoionization
  processes of many-electron atoms}},}\ }\href@noop {} {\bibfield  {journal}
  {\bibinfo  {journal} {Eur. Phys. J. Special Topics}\ }\textbf {\bibinfo
  {volume} {223}},\ \bibinfo {pages} {177} (\bibinfo {year}
  {2014})}\BibitemShut {NoStop}%
\bibitem [{\citenamefont {van~der Hart}\ and\ \citenamefont
  {Morgan}(2014)}]{Hart2014_Ne+}%
  \BibitemOpen
  \bibfield  {author} {\bibinfo {author} {\bibfnamefont {H.~W.}\ \bibnamefont
  {van~der Hart}}\ and\ \bibinfo {author} {\bibfnamefont {R.}~\bibnamefont
  {Morgan}},\ }\bibfield  {title} {\enquote {\bibinfo {title} {Population
  trapping in bound states during {IR}-assisted ultrafast photoionization of
  {Ne$^+$}},}\ }\href@noop {} {\bibfield  {journal} {\bibinfo  {journal} {Phys.
  Rev. A}\ }\textbf {\bibinfo {volume} {90}},\ \bibinfo {pages} {013424}
  (\bibinfo {year} {2014})}\BibitemShut {NoStop}%
\bibitem [{\citenamefont {Corkum}(1993)}]{Corkum1993}%
  \BibitemOpen
  \bibfield  {author} {\bibinfo {author} {\bibfnamefont {P.~B.}\ \bibnamefont
  {Corkum}},\ }\bibfield  {title} {\enquote {\bibinfo {title} {{Plasma
  perspective on strong field multiphoton ionization}},}\ }\href@noop {}
  {\bibfield  {journal} {\bibinfo  {journal} {Phys. Rev. Lett.}\ }\textbf
  {\bibinfo {volume} {71}},\ \bibinfo {pages} {1994} (\bibinfo {year}
  {1993})}\BibitemShut {NoStop}%
\bibitem [{\citenamefont {Shi}\ \emph {et~al.}(2015)\citenamefont {Shi},
  \citenamefont {Dong}, \citenamefont {Li}, \citenamefont {Wang},\ and\
  \citenamefont {Chen}}]{Shi2015}%
  \BibitemOpen
  \bibfield  {author} {\bibinfo {author} {\bibfnamefont {Y.~Z.}\ \bibnamefont
  {Shi}}, \bibinfo {author} {\bibfnamefont {F.~L.}\ \bibnamefont {Dong}},
  \bibinfo {author} {\bibfnamefont {Y.~P.}\ \bibnamefont {Li}}, \bibinfo
  {author} {\bibfnamefont {S.}~\bibnamefont {Wang}}, \ and\ \bibinfo {author}
  {\bibfnamefont {Y.~J.}\ \bibnamefont {Chen}},\ }\bibfield  {title} {\enquote
  {\bibinfo {title} {{Classical effect for enhanced high harmonic yield in
  ultrashort laser pulses with a moderate laser intensity}},}\ }\href@noop {}
  {\bibfield  {journal} {\bibinfo  {journal} {arXiv:1509.04457v1}\ } (\bibinfo
  {year} {2015})}\BibitemShut {NoStop}%
\bibitem [{\citenamefont {Yang}\ \emph {et~al.}(1993)\citenamefont {Yang},
  \citenamefont {Schafer}, \citenamefont {Walker}, \citenamefont {Kulander},
  \citenamefont {Agostini},\ and\ \citenamefont {DiMauro}}]{Yang1993}%
  \BibitemOpen
  \bibfield  {author} {\bibinfo {author} {\bibfnamefont {B.}~\bibnamefont
  {Yang}}, \bibinfo {author} {\bibfnamefont {K.~J.}\ \bibnamefont {Schafer}},
  \bibinfo {author} {\bibfnamefont {B.}~\bibnamefont {Walker}}, \bibinfo
  {author} {\bibfnamefont {K.~C.}\ \bibnamefont {Kulander}}, \bibinfo {author}
  {\bibfnamefont {P.}~\bibnamefont {Agostini}}, \ and\ \bibinfo {author}
  {\bibfnamefont {L.~F.}\ \bibnamefont {DiMauro}},\ }\bibfield  {title}
  {\enquote {\bibinfo {title} {{Intensity-dependent scattering rings in high
  order above-threshold ionization}},}\ }\href@noop {} {\bibfield  {journal}
  {\bibinfo  {journal} {Phys. Rev. Lett.}\ }\textbf {\bibinfo {volume} {71}},\
  \bibinfo {pages} {3770} (\bibinfo {year} {1993})}\BibitemShut {NoStop}%
\bibitem [{\citenamefont {Paulus}\ \emph {et~al.}(1994)\citenamefont {Paulus},
  \citenamefont {Becker}, \citenamefont {Nicklich},\ and\ \citenamefont
  {Walther}}]{Paulus1994}%
  \BibitemOpen
  \bibfield  {author} {\bibinfo {author} {\bibfnamefont {G.~G.}\ \bibnamefont
  {Paulus}}, \bibinfo {author} {\bibfnamefont {W.}~\bibnamefont {Becker}},
  \bibinfo {author} {\bibfnamefont {W.}~\bibnamefont {Nicklich}}, \ and\
  \bibinfo {author} {\bibfnamefont {H.}~\bibnamefont {Walther}},\ }\bibfield
  {title} {\enquote {\bibinfo {title} {{Rescattering effects in above-threshold
  ionization: a classical model}},}\ }\href@noop {} {\bibfield  {journal}
  {\bibinfo  {journal} {J. Phys. B}\ }\textbf {\bibinfo {volume} {27}},\
  \bibinfo {pages} {L703} (\bibinfo {year} {1994})}\BibitemShut {NoStop}%
\bibitem [{\citenamefont {Madsen}(2002)}]{Madsen2002}%
  \BibitemOpen
  \bibfield  {author} {\bibinfo {author} {\bibfnamefont {L.~B.}\ \bibnamefont
  {Madsen}},\ }\bibfield  {title} {\enquote {\bibinfo {title} {{Gauge
  invariance in the interaction between atoms and few-cycle laser pulses}},}\
  }\href@noop {} {\bibfield  {journal} {\bibinfo  {journal} {Phys. Rev. A}\
  }\textbf {\bibinfo {volume} {65}},\ \bibinfo {pages} {053417} (\bibinfo
  {year} {2002})}\BibitemShut {NoStop}%
\bibitem [{\citenamefont {Gavrila}(2002)}]{Gavrila2002}%
  \BibitemOpen
  \bibfield  {author} {\bibinfo {author} {\bibfnamefont {M.}~\bibnamefont
  {Gavrila}},\ }\bibfield  {title} {\enquote {\bibinfo {title} {{Atomic
  stabilization in superintense laser fields}},}\ }\href@noop {} {\bibfield
  {journal} {\bibinfo  {journal} {J. Phys. B}\ }\textbf {\bibinfo {volume}
  {35}},\ \bibinfo {pages} {R147} (\bibinfo {year} {2002})}\BibitemShut
  {NoStop}%
\bibitem [{Note2()}]{Note2}%
  \BibitemOpen
  \bibinfo {note} {To see a typical behavior of the phase gradient, for
  simplicity, let us consider the time evolution of a one-dimensional free
  electron prepared at $t=0$ in a normalized Gaussian wave packet having width
  $\Delta x$ and central momentum $k_0$: \begin {eqnarray*} &&\psi (x,t) =
  \protect \sqrt {\protect \frac {1}{\protect \sqrt {2\pi }}\protect \frac
  {\Delta x}{(\Delta x)^2+it/2}} \\ && \protect \hspace {12mm} \times \protect
  \qopname \relax o{exp}{\setbox \z@ \hbox {\frozen@everymath \@emptytoks
  \mathsurround \z@ $\nulldelimiterspace \z@ \left [\vcenter to\tw@ \big@size
  {}\right .$}\box \z@ } ik_0x -ik_0^2t/2 -\protect \frac
  {(x-k_0t)^2}{4[(\Delta x)^2+it/2]} {\setbox \z@ \hbox {\frozen@everymath
  \@emptytoks \mathsurround \z@ $\nulldelimiterspace \z@ \left ]\vcenter to\tw@
  \big@size {}\right .$}\box \z@ }. \end {eqnarray*} The derivative of the
  phase with respect to $x$, $d{\setbox \z@ \hbox {\frozen@everymath
  \@emptytoks \mathsurround \z@ $\nulldelimiterspace \z@ \left [\vcenter to\@ne
  \big@size {}\right .$}\box \z@ }\protect \qopname \relax o{arg}\psi
  (x,t){\setbox \z@ \hbox {\frozen@everymath \@emptytoks \mathsurround \z@
  $\nulldelimiterspace \z@ \left ]\vcenter to\@ne \big@size {}\right .$}\box
  \z@ }/dx=k_0+t(x-k_0t)/{\setbox \z@ \hbox {\frozen@everymath \@emptytoks
  \mathsurround \z@ $\nulldelimiterspace \z@ \left [\vcenter to\@ne \big@size
  {}\right .$}\box \z@ }4(\Delta x)^4+t^2{\setbox \z@ \hbox {\frozen@everymath
  \@emptytoks \mathsurround \z@ $\nulldelimiterspace \z@ \left ]\vcenter to\@ne
  \big@size {}\right .$}\box \z@ }$, linearly increases with $x$, and is
  greater than $k_0$ for $x>k_0t$. Also see the discussion in Ref.~\cite
  {Sidky2000}.}\BibitemShut {Stop}%
\bibitem [{\citenamefont {Gallopoulos}\ and\ \citenamefont
  {Saad}(1992)}]{Gallopoulos1992}%
  \BibitemOpen
  \bibfield  {author} {\bibinfo {author} {\bibfnamefont {E.}~\bibnamefont
  {Gallopoulos}}\ and\ \bibinfo {author} {\bibfnamefont {Y.}~\bibnamefont
  {Saad}},\ }\bibfield  {title} {\enquote {\bibinfo {title} {{Efficient
  solution of parabolic equations by Krylov approximation methods}},}\
  }\href@noop {} {\bibfield  {journal} {\bibinfo  {journal} {SIAM J. Sci.
  Statist. Comput.}\ }\textbf {\bibinfo {volume} {13}},\ \bibinfo {pages}
  {1236} (\bibinfo {year} {1992})}\BibitemShut {NoStop}%
\bibitem [{\citenamefont {Saad}(1992)}]{Saad1992}%
  \BibitemOpen
  \bibfield  {author} {\bibinfo {author} {\bibfnamefont {Y.}~\bibnamefont
  {Saad}},\ }\bibfield  {title} {\enquote {\bibinfo {title} {{Analysis of some
  Krylov subspace approximations to the matrix exponential operator}},}\
  }\href@noop {} {\bibfield  {journal} {\bibinfo  {journal} {SIAM J. Numer.
  Anal.}\ }\textbf {\bibinfo {volume} {29}},\ \bibinfo {pages} {209} (\bibinfo
  {year} {1992})}\BibitemShut {NoStop}%
\end{thebibliography}
\end{document}